\documentclass[11pt, a4paper, onecolumn, copyright, google]{google}

\uselogo{}

\usepackage{kantlipsum, lipsum}
\usepackage{dm-colors}
\usepackage{amsmath}
\usepackage{pstricks, pst-node}
\usepackage{verbatim}
\usepackage{multirow}
\usepackage{makecell}
\usepackage{scalerel}
\usepackage{booktabs}
\usepackage{enumitem}
\usepackage{xspace}
\usepackage{bm}
\usepackage{bbm}
\usepackage{mathtools}
\usepackage{soul}
\usepackage{epsfig}
\usepackage{graphicx}
\usepackage{tcolorbox}
\usepackage{subcaption}
\usepackage{amssymb}
\usepackage{colortbl}
\usepackage{csquotes}
\usepackage{etex}
\usepackage{setspace}
\usepackage{svg}
\usepackage{colortbl}
\usepackage{tabularx,ragged2e}
\usepackage{placeins}
\usepackage{afterpage}
\usepackage[symbol]{footmisc}
\usepackage[bibstyle=nature,citestyle=numeric-comp,%
           natbib=true,backend=biber,maxbibnames=99,%
           giveninits=false,sorting=none]{biblatex}
\usepackage{nameref}
\usepackage{varioref}
\hypersetup{
    breaklinks=false, 
    colorlinks=true,
    citecolor=ourdarkblue,
    urlcolor=ourdarkblue,
    linkcolor=ourdarkblue
}
\usepackage[noabbrev,capitalize]{cleveref}

\definecolor{headerblue}{RGB}{30, 64, 110}
\definecolor{subheader}{RGB}{70, 110, 160}
\definecolor{airow}{RGB}{232, 240, 252}
\definecolor{humanrow}{RGB}{240, 252, 244}
\definecolor{passgreen}{RGB}{34, 139, 34}
\definecolor{warnred}{RGB}{180, 30, 30}

\definecolor{tableheader}{RGB}{30, 64, 110}
\definecolor{tablesubheader}{RGB}{70, 110, 160}
\definecolor{rowaltlight}{RGB}{240, 245, 252}
\definecolor{highlightgold}{RGB}{255, 243, 205}

\usepackage{etoc}
\usepackage{lineno}
\graphicspath{{figures/}}

\usepackage{listings}
\DeclareFixedFont{\ttb}{T1}{txtt}{bx}{n}{9} 
\DeclareFixedFont{\ttm}{T1}{txtt}{m}{n}{9}  
\usepackage{color}
\definecolor{deepblue}{rgb}{0,0,0.5}
\definecolor{deepred}{rgb}{0.6,0,0}
\definecolor{deepgreen}{rgb}{0,0.5,0}
\newcommand\pythonstyle{\lstset{
language=Python,
basicstyle=\footnotesize,
morekeywords={self},              
keywordstyle=\color{deepblue},
emph={},          
emphstyle=\color{deepred},    
stringstyle=\color{deepgreen},
commentstyle=\color{deepgreen},
showstringspaces=false
}}
\lstnewenvironment{python}[1][]
{
\pythonstyle
\lstset{#1}
}
{}

\newcommand\pythoninline[1]{{\pythonstyle\lstinline!#1!}}

\newtcolorbox{promptbox2}[1]{ 
    colback=black!5!white,
    colframe=black!60!white,
    arc=1mm, 
    boxrule=1pt, 
    bottomrule=2pt,
    title=#1, 
    fonttitle=\bfseries,
    fontupper=\fontfamily{lmtt},
    fontlower=\fontfamily{lmtt},
    capture=minipage 
}

\title{A prospective clinical feasibility study\\of a conversational diagnostic AI in\\an ambulatory primary care clinic}

\renewcommand{\today}{2026-03-09}

\author[$\ast$,3]{Peter Brodeur}
\author[$\ast$,3]{Jacob M. Koshy}
\author[$\ast$,1]{Anil Palepu}
\author[2]{Khaled Saab}
\author[3]{Ava Homiar}
\author[1]{Roma Ruparel}
\author[3]{Charles Wu}
\author[2]{Ryutaro Tanno}
\author[1]{Joseph Xu}
\author[1]{Amy Wang}
\author[2]{David Stutz}
\author[2]{Wei-Hung Weng}
\author[3]{Hannah M. Ferrera}
\author[2]{David Barrett}
\author[3]{Lindsey Crowley}
\author[1]{Jihyeon Lee}
\author[4]{Spencer E. Rittner}
\author[1]{Ellery Wulczyn}
\author[5]{Selena K. Zhang}
\author[2]{Elahe Vedadi}
\author[6]{Christine G. Kohn}
\author[1]{Kavita Kulkarni}
\author[3]{Vinay Kadiyala}
\author[2]{S. Sara Mahdavi}
\author[3]{Wendy Du}
\author[1]{Jessica M. Williams}
\author[3]{David Feinbloom}
\author[1]{Renee Wong}
\author[2]{Tao Tu}
\author[1]{Petar Sirkovic}
\author[1]{Alessio Orlandi}
\author[1]{Christopher Semturs}
\author[1]{Yun Liu}
\author[1]{Juraj Gottweis}
\author[1]{Dale R. Webster}
\author[2]{Joëlle Barral}
\author[1]{Katherine Chou}
\author[2]{Pushmeet Kohli}
\author[1]{Avinatan Hassidim}
\author[1]{Yossi Matias}
\author[1]{James Manyika}
\author[4]{Rob Fields}
\author[3]{Jonathan X. Li}
\author[$\dagger$,3]{Marc L. Cohen}
\author[$\dagger$,2]{Vivek Natarajan}
\author[$\dagger$,1]{\\Mike Schaekermann}
\author[$\dagger$,2]{Alan Karthikesalingam}
\author[$\dagger$,1,3]{Adam Rodman}

\affil[*]{Equal contributions}
\affil[$\dagger$]{Equal leadership}

\affil[1]{Google Research}
\affil[2]{Google DeepMind}
\affil[3]{Beth Israel Deaconess Medical Center}
\affil[4]{Beth Israel Lahey Health}
\affil[5]{Harvard Medical School}
\affil[6]{Massachusetts General Hospital}

\correspondingauthor{$\{$pbrodeur$,\,$jkoshy$,\,$arodman$\}$@bidmc.harvard.edu, $\{$mikeshake$,\,$alankarthi$\}$@google.com}

\begin{document}

\begin{refsection}

\begin{abstract}
Large language model (LLM)-based AI systems have shown promise for patient-facing diagnostic and management conversations in simulated settings. Translating these systems into clinical practice requires assessment in real-world workflows with rigorous safety oversight. We report a prospective, single-arm feasibility study of an LLM-based conversational AI, the Articulate Medical Intelligence Explorer (AMIE), conducting clinical history taking and presentation of potential diagnoses for patients to discuss with their provider at urgent care appointments at a leading academic medical center. 100 adult patients completed an AMIE text-chat interaction up to 5 days before their appointment. We sought to assess the conversational safety and quality, patient and clinician experience, and clinical reasoning capabilities compared to primary care providers (PCPs). Human safety supervisors monitored all patient-AMIE interactions in real time and did not need to intervene to stop any consultations based on pre-defined criteria. Patients reported high satisfaction and their attitudes towards AI improved after interacting with AMIE (p < 0.001). PCPs found AMIE's output useful with a positive impact on preparedness. AMIE's differential diagnosis (DDx) included the final diagnosis, per chart review 8 weeks post-encounter, in 90\% of cases, with 75\% top-3 accuracy. Blinded assessment of AMIE and PCP DDx and management (Mx) plans suggested similar overall DDx and Mx plan quality, without significant differences for DDx (p = 0.6) and appropriateness and safety of Mx (p = 0.1 and 1.0, respectively). PCPs outperformed AMIE in the practicality (p = 0.003) and cost effectiveness (p = 0.004) of Mx. While further research is needed, this study demonstrates the initial feasibility, safety, and user acceptance of conversational AI in a real-world setting, representing crucial steps towards clinical translation.
\end{abstract}

\maketitle


\afterpage{
\begin{figure}[ht!]
    \centering
    \includegraphics[width=1\textwidth,height=\textheight,keepaspectratio]{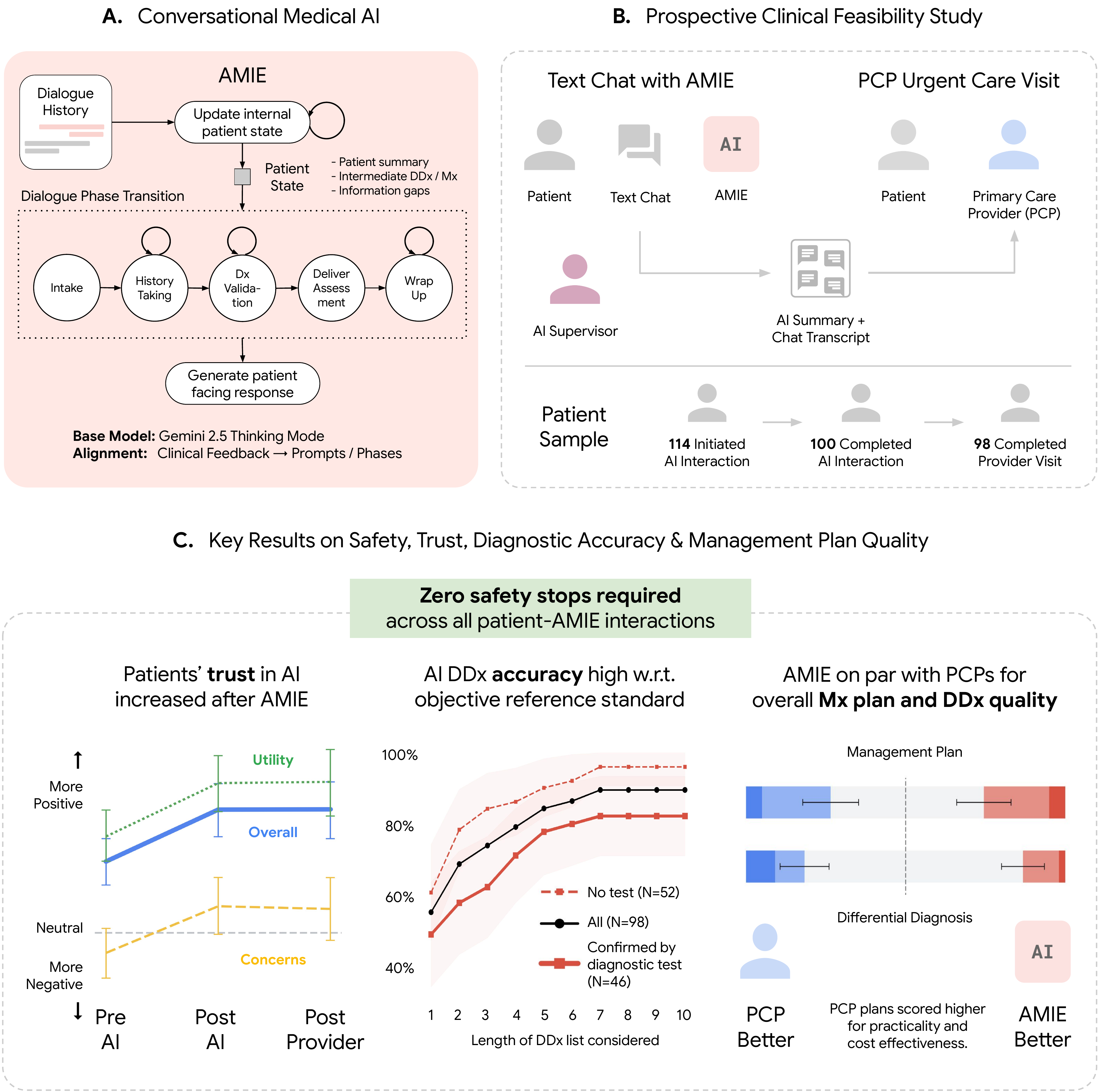}
    \caption{\textbf{Overview of contributions.} We adapted the AMIE system described in \citet{saab2025advancing} \textbf{(A)} to conduct a first-of-its-kind prospective clinical feasibility study of AMIE in an ambulatory primary care clinic \textbf{(B)}, producing evidence regarding the safety, feasibility, conversation quality, and clinical reasoning performance of AMIE, as well as patient and provider experience with the system \textbf{(C)}.}
    \label{fig:contributions}
\end{figure}
}

\section{Introduction}
\label{sec:introduction}
There is an ever-worsening shortage of primary care physicians (PCPs) affecting virtually every country in the world \cite{Walensky2025-ld, Adashi2025-jr, Lawson2023-gf}. 
Tasked with a greater workload and an aging population, burnout rates have surged among PCPs\cite{Abraham2020-mb}. 
Technology that facilitates effective use of electronic health records (EHRs) and improves medical team efficiency holds promise for improving accessibility of care and reducing physician burnout\cite{Rotenstein2024-ah, Olson2025}, while digital care pathways involving telehealth encounters and artificial intelligence (AI) systems for smart intake have become increasingly popular \cite{zeltzer2023diagnostic, Chang2024-kp}.
Large language models (LLMs) show particular promise for extending care availability, with capabilities to engage in nuanced clinical reasoning and conversation \cite{nori2025sequentialdiagnosislanguagemodels, brodeur2025superhumanperformancelargelanguage, buckley2025advancingmedicalartificialintelligence}.
Real-world deployments of patient-facing conversational AI further indicate that such systems can meaningfully contribute to patient care coordination and other intake tasks \cite{Zeltzer2025-qi, Tao2026-nl}, though they have not yet been extensively tested for clinical settings with real human-in-the-loop workflows with PCPs.

Our previous work introduced the Articulate Medical Intelligence Explorer (AMIE), an LLM-based system optimized for clinical dialogue \cite{saab2025advancingconversationaldiagnosticai, palepu2025conversationalaidiseasemanagement, Tu2025-lf,Vedadi2025}.
Through studies simulating Objective Structured Clinical Examinations (OSCEs) with trained patient actors, AMIE demonstrated proficiency that was comparable, and in some aspects superior, to human PCPs in diagnostic reasoning during simulations of initial encounters, longitudinal disease management across multiple visits, and encounters requiring clinical reasoning over multimodal artifacts of care \cite{Tu2025-lf, saab2025advancingconversationaldiagnosticai, palepu2025conversationalaidiseasemanagement}.
Despite these promising results in simulated consultations, the safe and effective translation of such AI systems into real-world clinical practice remains underexplored.
The capability to perform a high-quality diagnostic conversation is just one component of safe deployment in care delivery. For AI systems to be effective, they must also adhere to strict safety criteria, especially since emerging evidence suggests that LLM care plans have the potential to cause harm, undertriage medical concerns or unreliably address mental health issues, and that patients are not always able to distinguish between inadequate and adequate medical advice \cite{wu2025firstnoharmclinicallysafe, Ramaswamy2026, Shekar2025-bt}.

Real-world patient interactions introduce complexities not fully captured by standardized actors or scenarios, including diverse communication styles, unpredictable clinical presentations, a spectrum of health and technology literacy, emotions such as anxiety, and varying levels of management urgency \cite{Mancia2021-kn}.
Collectively, these real-world complexities raise the bar for building conversational AI that remains consistently and reliably safe. Furthermore, assessing the integration of AI tools into existing clinical workflows and gauging the perspectives of both patients and clinicians are vital steps for successful adoption.

To bridge the gap between simulated evaluations and clinical application, we conducted a prospective, single-arm feasibility study of AMIE performing pre-visit clinical conversations within a real-world clinical setting.
AMIE collected a detailed history directly from real patients seeking urgent care and presented them information related to possible diagnoses to prepare them for their appointment. This information was then delivered to their PCPs prior to their appointments.
Given the high stakes nature of this workflow, our primary objectives were to evaluate the safety of AMIE engaging in conversational history taking with actual patients scheduled for urgent care appointments at a primary care clinic within a leading, high-volume academic medical center, as well as the overall quality of these conversations and operational factors required for successful study completion.

Secondary objectives focused on assessing patient and provider perceptions of the interaction from a qualitative perspective through semi-structured interviews, and the quality of AI-generated outputs, including a list of potential diagnoses presented in the interaction transcript and management plans which were not shown to patients or providers, but logged for research purposes only. Using an EHR for chart review, we reviewed eight weeks of clinical data and assessments to identify the ``ground truth'' final diagnosis of presenting complaints, thereby enabling an evaluation of the quality of the diagnostic and management plan generated by AI in comparison to PCPs based on information from this initial urgent care appointment.
This study represents a necessary step in understanding the practicalities and challenges of deploying advanced conversational AI in patient care workflows.

Our contributions are summarized in \cref{fig:contributions}:

\begin{itemize}[leftmargin=1.5em,rightmargin=0em]
\setlength\itemsep{5pt}

\item We designed an AMIE system inspired by previous work \cite{saab2025advancing} for this real-world clinical study setting, leveraging the more recent family of Gemini 2.5 base models \cite{comanici2025gemini} with Thinking Mode enabled, alongside a refined state-aware chain-of-reasoning strategy that was adapted for robust pre-visit conversational clinical history-taking with patients through iterations of clinical testing and feedback (\cref{fig:contributions}.a).

\item Using this adapted version of AMIE, we conducted a pre-registered prospective feasibility study evaluating AMIE interacting with 100 patients in a real-world clinical workflow embedded within an ambulatory primary care clinic.
Based on the information shared during the AI encounter, AMIE presented information about potential diagnoses and next steps that their clinician may want to discuss with them.
The conversation transcript, along with an automatically generated summary, was provided to the PCP prior to the urgent care appointment.
We also developed a safety oversight protocol for all AMIE encounters with prespecified criteria for safety interruptions from supervising physicians (\cref{fig:contributions}.b).

\clearpage
\item We provide results on the system's conversational safety, feasibility, conversation quality, patient and provider perceptions, and preliminary clinical reasoning performance including diagnostic accuracy and management plan quality based on chart review.
We observed zero safety stops across all patient-AMIE interactions and patient attitudes towards AI, measured using the General Attitudes towards AI Scale (GAAIS), increased after interacting with AMIE (p < 0.001).
AMIE’s differential diagnosis (DDx) included the final diagnosis, per chart review 8 weeks post-encounter, in 90\% of cases, with a top-3 accuracy of 75\%.
Blinded assessment of AMIE and PCP DDx and management (Mx) plans suggested similar overall DDx and Mx plan quality, without significant differences for DDx (p = 0.6) and appropriateness and safety of the Mx plan (p = 0.1 and 1.0, respectively). PCPs outperformed AMIE in the practicality (p = 0.003) and cost effectiveness (p = 0.004) of Mx plans (\cref{fig:contributions}.c)
\end{itemize}

\section{AI System}
\label{sec:system}

AMIE is a conversational diagnostic AI system designed to interact with patients in a synchronous text-based interface \cite{Tu2025-lf}.
The system was built upon Gemini 2.5 Pro (knowledge cutoff January 2025) without any domain-specific fine-tuning.
For the purpose of this clinical study, we used the agent setup described in \citet{saab2025advancing} as a starting point, using the Gemini 2.5 family of models \cite{comanici2025gemini} with Thinking mode enabled, and further aligned the prompts and conversation phases of this agentic system to adapt it to the specific study setting of AI-driven clinical history-taking prior to an real-world urgent care appointment.
This alignment was done based on extensive feedback from clinical experts, as well as synthetic multi-turn roll-outs of dialogues with AI-simulated patients similar to the process described in previous studies \cite{saab2024capabilities,palepu2025conversationalaidiseasemanagement}.
Through this process, we developed an agent which continuously maintains a rich internal state---including an up-to-date patient summary, working differential diagnosis, pertinent information gaps, and draft management plan---to inform its reasoning. Upon meeting a new patient, the system is designed to proactively guide the patient interaction through five distinct phases, with unique prompting and internal logic for each as (\cref{fig:contributions}.a):

\begin{itemize}
\item \textbf{Intake.} The agent initiates the consultation, establishing rapport and eliciting basic demographics and the patient’s chief complaint.
\item \textbf{History Taking.} The agent conducts adaptive inquiry to comprehensively understand the patient's symptoms and relevant history. 
Rather than following a static script, questions are dynamically generated based on the agent's diagnostic hypotheses and information gaps.
\item \textbf{Diagnostic Validation.} Anchoring on its provisional hypotheses, the agent seeks to fully characterize the patient's condition, improve its confidence, disambiguate the differential diagnosis, and revise hypotheses as needed, prior to sharing its assessment.
Before proceeding, the agent also summarizes its understanding to the patient, inviting corrections or clarifications to ensure data accuracy.
\item \textbf{Deliver Assessment.} The agent presents possible diagnosis and management options to consider.
Given the real-world context of this study, these outputs are framed tentatively---accompanied by clear disclaimers---as possible diagnoses or next steps for the patient to discuss with a provider.
\item \textbf{Consultation Wrap-up.} The agent confirms the patient's understanding and provides them space to continue asking questions.
It continues to clarify and, in certain cases, update its assessment until the patient elects to conclude the encounter.
\end{itemize}

After completing a consultation, AMIE generates a conversation summary, which, along with the conversation transcript, was shared with PCPs in our study.
For research purposes only, AMIE also generated a management plan, which was not directly shared with the patient or provider, but logged for the purpose of evaluating AMIE.

\section{Methods}\label{sec:methods}

\begin{figure}[t!]
    \includegraphics[width=\textwidth,height=\textheight,keepaspectratio]{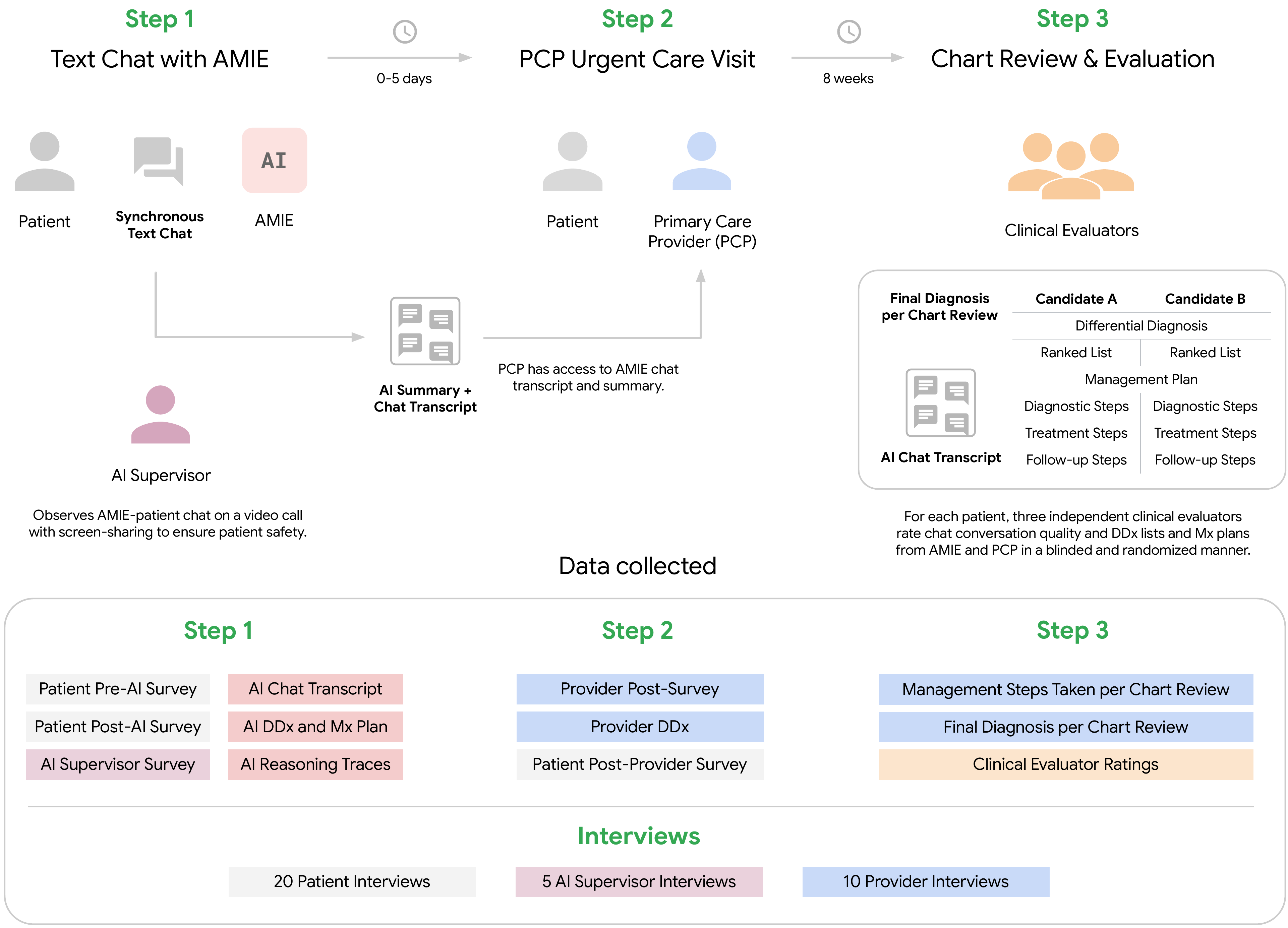}
    \vspace{0.5em}
    \caption{\textbf{Study design.} For each patient, the study flow consisted of three distinct steps: \textbf{(1)} The patient conversed with AMIE through a synchronous text chat interface, while an AI supervisor observed the patient-AMIE chat on a video-call with screen-sharing to ensure patient safety; \textbf{(2)} the patient saw a primary care physician (PCP) in person or via telehealth up to five days after the AMIE chat, with the PCP having access to the AMIE chat transcript and summary; \textbf{(3)} eight weeks after the provider visit, the final diagnosis was extracted from the patient's chart and three independent clinical evaluators rated the chat conversation quality as well as the differential diagnosis and management plan from both AMIE and PCP in a blinded and randomized manner.}
    \label{fig:study_design}
\end{figure}

An overview of our study design including the data collected in the study is provided in \cref{fig:study_design}.

\subsection{Study Setting and Oversight}

We conducted a prospective, single-arm feasibility study to evaluate AMIE's ability to conduct a real-world pre-visit clinical conversation. 
The study was performed at Healthcare Associates (HCA), part of Beth Israel Deaconess Medical Center (BIDMC), in Boston, MA from April 2025 to November 2025.
HCA is an academic primary care practice with 56 attending physicians, 110 resident physicians, 15 nurse practitioners, and approximately 40,000 total patients.

The study protocol was approved by the BIDMC Committee on Clinical Investigations (FWA00003245, IRB protocol 2024P000095) and pre-registered on ClinicalTrials.gov (NCT06911398).
All patient participants provided written informed consent and HIPAA authorization electronically via REDCap prior to participation.
REDCap (Research Electronic Data Capture) is a secure, web-based software platform designed to support data capture for research studies, providing 1) an intuitive interface for validated data capture; 2) audit trails for tracking data manipulation and export procedures; 3) automated export procedures for seamless data downloads to common statistical packages; and 4) procedures for data integration and interoperability with external sources \cite{harris2009redcap,harris2019redcap}.
All PCP participants received a study information sheet as part of the consent procedure.
This study is reported in accordance with the guidelines set forth by TRIPOD-LLM \cite{Gallifant2025} and the corresponding checklist is provided in \cref{appendix:tab:tripod_llm_checklist}.

\subsection{Patient Eligibility}

Eligible patient participants were scheduled for either an in-person or telehealth urgent care visit at HCA with a single chief complaint and had already been determined by clinic triage staff not to need emergency care.
They must also have been established patients at HCA, aged 18 years or older, with English listed as their primary language in the electronic health record (EHR), be actively enrolled in HCA's online patient portal, and be able to interact with AMIE remotely using a personal device other than a cell phone, such as a laptop or desktop computer.
Due to institutional IRB constraints, we excluded patients with known pregnancy. We also excluded mental health-related chief complaints as the AMIE system has not been validated for psychiatric concerns in prior simulated studies.

\subsection{Patient and PCP Recruitment}

Patients were screened daily during the study period by a research coordinator.
If determined to be eligible, potential participants then received an electronic study information sheet via the HCA's online patient portal.
This electronic recruitment clearly stated the risks of participation and that their decision to participate would not affect their care or access to care.
Each patient was offered 25 US dollars to participate in the encounter with AMIE, and an additional 25 US dollars for an optional post-AMIE interaction user experience interview.
Due to slower than anticipated recruitment, midway through the study period, these incentives were increased to 50 US dollars.
After receiving the electronic recruitment invitation, patients were contacted by phone to participate.
Participants signed an electronic consent form, which was also signed by the research assistant and recorded in a secure REDCap database.
Given the goal of a feasibility study, recruitment was prespecified to continue until 200 patients were consented, or 100 full conversations were completed, whichever came first.
The total number of patients was selected based on clinical feasibility given the patient volume at HCA, and to be comparable to prior AMIE evaluations with patient actors \cite{Tu2025-lf}.

PCPs were consented either in person or electronically from a subset of providers (attending physicians, nurse practitioners, and internal medicine residents) who saw urgent care patients as part of their routine practice.
For electronic consent, PCPs received an information sheet that clearly outlined the study purpose, objectives, safety criteria, and compensation.
Residents were also included in the study, though their participation was contingent on the final plan for the urgent care visit being developed with attending physician oversight.
Nurse practitioners in this clinic provide care without direct oversight from attending physicians.
While eligible patients were established patients at the clinic with a longitudinal PCP, the PCP managing the urgent care appointment was not necessarily the patient's longitudinal PCP.
For example, if the patient's longitudinal PCP was unavailable the day the patient was presenting, they would be seen by any PCP with an open schedule.

\subsection{Intervention Protocol}

Following informed consent, patients were scheduled for a remote interaction with AMIE, conducted 0-5 days prior to their planned urgent care visit.
Prior to interacting with AMIE, patients completed a brief pre-interaction REDCap survey that captured baseline demographics (age, gender identity, race/ethnicity), health literacy, technology literacy, prior use of chatbots, and the GAAIS (\cref{tab:participation}, B; \cref{appendix:fig:patient_pre_amie_survey}) \cite{Schepman2020-ih}. 
Patients remotely accessed AMIE through a secure synchronous text chat interface. 
During the interaction, AMIE's primary task was to conduct a clinical conversation focused on the patient's presenting complaint.
Concluding the conversation, AMIE also presented information about potential diagnoses to the patient and, if requested, next steps the provider may want to discuss with the patient during the PCP visit.

Each patient-AMIE interaction was continuously monitored in real-time by a trained, board-certified internal medicine physician from BIDMC, designated as the AI supervisor, via a secure video call with screen sharing from a designated study account. The AI supervisor was not visible to the patient once the patient-AMIE interaction commenced.
The AI supervisor's role was to ensure patient safety.
Seven board certified internal medicine physicians served as supervisors throughout the study.
Through explicit safety training, they were instructed to intervene and interrupt the chat if it met any of the following pre-defined stop criteria:
(1) immediate concern for harm to self or others,
(2) significant emotional distress exhibited by the patient related to the AI interaction,
(3) potential for clinical harm identified by the supervisor based on the conversation, or
(4) an explicit request from the patient to end the session.
These criteria and corresponding safety plans were reviewed with the supervisors as part of their training. 

Immediately after the chat concluded, the AI supervisor conducted a debrief session with the patient while still on the same video call.
During this debrief, the supervisor was instructed to address any concerns, clarify information, and correct any errors or AI hallucinations identified during the interaction.
Immediately following the conversation, the AI supervisor completed a brief REDCap survey on the presence and type of safety interrupts.

Upon completion of the chat conversation and AI supervisor debrief, the patient also completed a REDCap survey which captured a repeat GAAIS (also done in the pre-interaction REDCap survey), enabling comparisons of opinion shifts.
This survey was also meant to capture patient satisfaction and perceived conversation quality via the complete General Medical Counsel Patient Questionnaire (GMCPQ), the Practical Assessment of Clinical Examination Skills (PACES) components for ‘Managing patient concerns’ and  ‘Maintaining patient welfare’ and the Patient-Centered Communication Best Practices (PCCBP) rubric on relationship fostering \cite{Tu2025-lf} (\cref{appendix:fig:patient_post_amie_survey}).

Following these activities, the patient attended their urgent care visit as planned with the PCP they had been scheduled to see. This was never the same provider as the AI supervisor.
The patient-AMIE transcript, along with an automatically generated summary of the conversation, was provided to the PCP prior to the start of the urgent care visit for their review. This document included the list of potential diagnoses but not the AMIE management plan which was stored separately for comparative analyses only. Because of resident workflow in HCA where residents can staff with multiple attendings unpredictably, in the case of a resident physician visit, the resident, not the attending, was sent the transcript.
After the visit, the patient completed one final REDCap survey assessing perceptions of visit efficiency and a final repeat of sentiments towards AI, as was captured in the previous two REDCap surveys (\cref{appendix:fig:patient_post_pcp_survey}).

After the urgent care visit, the PCP completed a REDCap survey to assess whether they had reviewed the patient-AMIE transcript and summary, and to evaluate their sentiments towards AMIE regarding preparedness for the visit, harm, trust, and behavior change (\cref{appendix:fig:pcp_post_visit_survey}).

\subsection{Safety Pilot}

Prior to launching the full study, a limited pilot involving ten patients was performed using five attending physician PCPs, followed by a prespecified safety pause.
During this safety pause, no further patient recruitment was done until study staff at BIDMC reviewed all study data, including chat transcripts, to ensure the effectiveness of safety supervision.
No changes were made to the study protocol after the safety pause.

\subsection{Semi-Structured Interviews}

During recruitment for the AMIE encounter, patients were asked whether they were willing to participate in an interview with a study team member and share their experiences with the chatbot. 
For patients willing to participate in the interview, during the initial phase of the study, research coordinators verified the time of their urgent care appointment and scheduled interviews to be conducted after both the AMIE encounter and urgent care appointment had concluded. However, due to high attrition, the research team later scheduled interviews with patients immediately after their AMIE encounters.

Patient participants who took part in interviews were asked to describe their experience using AMIE prior to their urgent care visit, and to explain how that conversation influenced their visit preparedness, understanding, and trust in the chatbot.
In addition, if they had completed the urgent care appointment, patients were asked to elaborate on how the chatbot affected the interaction with their clinician, compared with usual care, as well as their self-directed information-seeking behaviors (e.g., Google search). 

The research team later interviewed a subset of PCPs and all AI supervisors who participated in the study to collect information on how the chatbot affected clinician workflow, clinical preparedness, and patient encounters in real-world care settings.
These interviews posed questions to inform researchers on feasibility and future implementation decisions.
AI supervisors were prompted to evaluate the safety and supervisory burden of having a trained staff member observe patients interacting with AMIE in real time, particularly during early deployment.
They were also asked to identify technical and onboarding barriers affecting the patient-AMIE interactions. 

\subsection{Data Security}

Chat transcripts were temporarily stored in a secure cloud storage bucket at BIDMC.
Prior to analysis or sharing with external collaborators, a BIDMC team member de-identified all transcripts according to Safe Harbor criteria, manually removing all protected health information (PHI) \cite{Committee_on_Strategies_for_Responsible_Sharing_of_Clinical_Trial_Data2015-be}.
These were further manually audited by A.R. from the study team to verify PHI stripping prior to transfer to Google for analysis via a secure data bucket.

\subsection{Infrastructure Challenges}

After 50 encounters were completed, considerable latency occurred due to compute limitations on the research infrastructure used for this study. The system's base model was switched from Gemini 2.5 Pro to Gemini 2.5 Flash, successfully reducing latency with only minimal expected degradation in performance.

\subsection{Outcome Measures}

\subsubsection{Primary Outcomes}

The study's pre-registered primary outcomes focused on (a) the safety and feasibility of AMIE conversations, (b) the quality of AMIE's clinical dialogue, and (c) patient and PCP experiences with AMIE in this real-world setting.
Specifically, primary outcomes included:
(a) the total number and type of chat terminations as determined by the AI supervisors;
(b) the quality of AMIE's clinical dialogue assessed from the perspective of patients and clinical evaluators using a combination of GMCPQ, PACES and PCCBP rubrics from prior work \cite{Tu2025-lf};
(c) experience surveys collected from patients before and after the AMIE encounter, and from both patients and PCPs after the urgent care encounter (\cref{appendix:survey_details}).
Clinical evaluators consisted of a panel of eight board-certified internal medicine physicians (HF, LC, SR, CK, VK, WD, MC, JL) who reviewed AMIE's conversations and applied several standardized rubrics to assess their quality (\cref{appendix:clinical_evaluator_ratings}), including rubrics that are routinely used in the evaluation of medical trainees, and have been used in prior studies of patient-facing chatbots \cite{Tu2025-lf}.
There was overlap between the group of AI supervisors and the group of clinical evaluators, and we ensured that clinical evaluators were only assigned patient cases for quality review which they had not themselves overseen as an AI supervisor.

\begin{table}[t!]
    \centering
    \includegraphics[width=\textwidth,keepaspectratio]{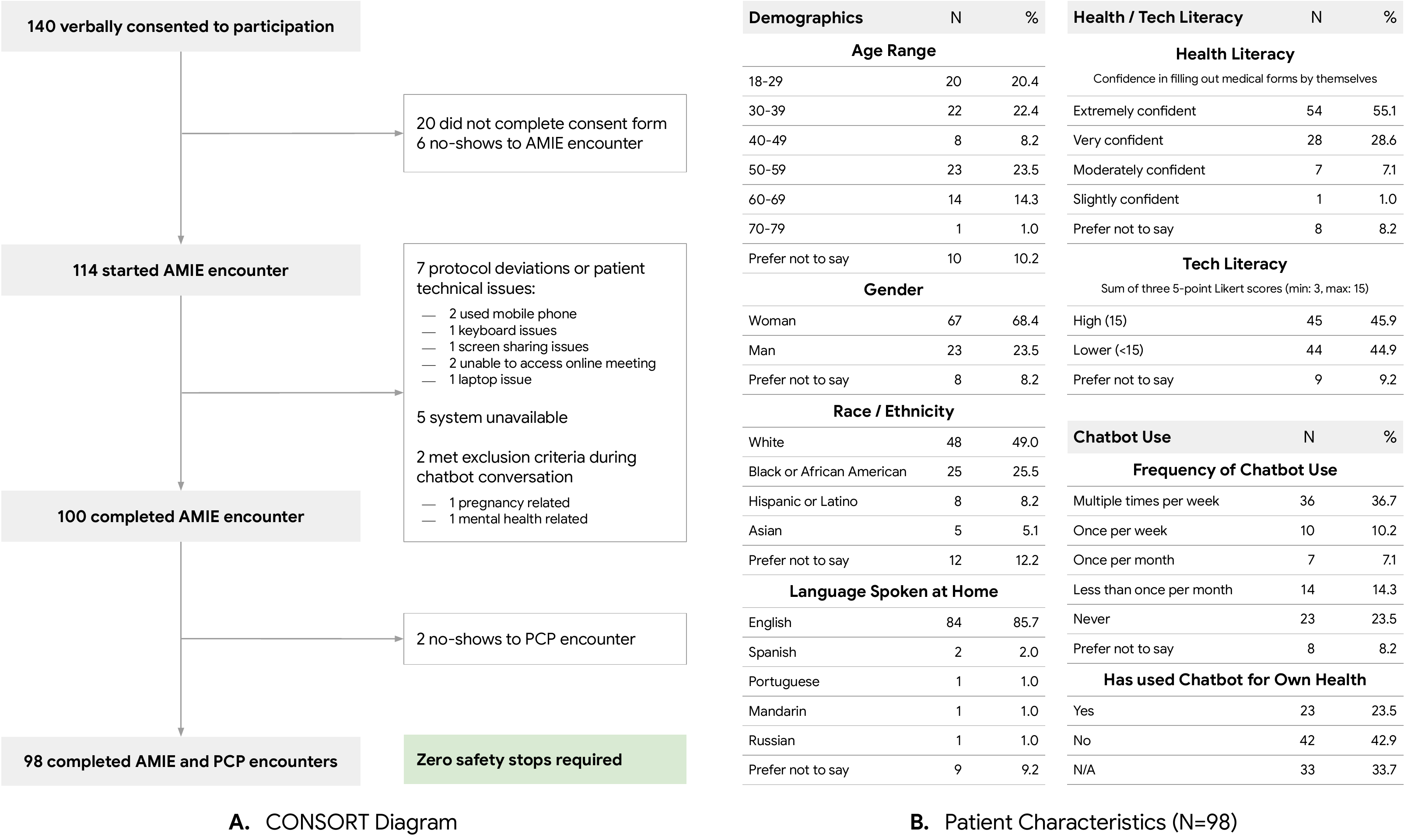}
    \vspace{0.5em}
    \caption{\textbf{Patient Participation Statistics.} The CONSORT diagram provides statistics regarding the flow of patient participants through our study \textbf{(A)}. For the 98 patients who completed both the AMIE encounter and the PCP encounter, we provide patient characteristics based on survey responses collected from patients before interacting with AMIE. \textbf{(B)}. Counts for \textit{`Prefer not to say'} and \textit{`N/A'} include the eight patients who did not complete the pre-AMIE survey.}
    \label{tab:participation}
\end{table}

\subsubsection{Secondary Outcomes}

Secondary outcomes focused on: (a) patient, AI supervisor, and PCP experiences as captured through semi-structured interviews; (b) clinical reasoning performance of AMIE and PCPs. Semi-structured interviews were used to triangulate survey responses from patients and PCPs to elucidate the underlying rationale for scale-based assessments. Clinical reasoning performance were assessed in two ways: AMIE's diagnostic accuracy, and quality of management plans and differential diagnoses of both AMIE and PCPs.

First, diagnostic accuracy for AMIE was assessed relative to the ``ground truth'' final diagnosis for each patient case.
For each case, the final diagnosis was established through a retrospective chart review conducted eight weeks post-visit by a panel of three internists (AR, JK, PB) blinded to AMIE’s output.
The panel reviewed each patient's longitudinal EHR, including follow-up laboratory results, imaging, and specialist notes, to determine the most reasonable final diagnosis.
AMIE's diagnostic accuracy was then assessed via two board-certified internal medicine physicians (AR, JK) reviewing all of AMIE's differential diagnoses, chat transcripts and the final diagnosis and rating of AMIE's differential diagnoses on the Bond/Graber scale \cite{bond2012differential}.
The Bond/Graber scale is a 5-point scale of differential quality.
For scores of 5 (\textit{`The actual diagnosis was suggested in the differential'}) and 4 (\textit{`The suggestions included something very close, but not exact'}), the raters also calculated the rank of the correct diagnosis within AMIE's differential diagnosis list for each case.
This process was done by each of the two raters separately, followed by resolution of rating discrepancies through deliberation among the two raters.

Second, the quality of management plans and differential diagnoses of both AMIE and PCPs was assessed by clinical evaluators in a blinded and randomized manner. In the study, only the list of potential diagnoses was shown to patients and PCPs via the patient-AMIE transcript. The management plan generated by AMIE was stored separately for research purposes only and not shown to either patients or PCPs.
Clinical evaluators were the same group of eight board-certified internal medicine physicians who also provided assessments of AMIE's conversation quality for primary outcome measures.
AMIE's differential diagnoses and management plans were logged by the system and stored for the purpose of this assessment.
For PCPs, differential diagnoses and management plans were extracted via chart review by the same panel who extracted the final diagnosis for each case. Management plans were sub-divided into diagnostic steps, treatment steps and follow-up steps.
To eliminate AI vs. PCP provenance bias during ratings, a blinding procedure was used to transform management plans and differential diagnosis from AMIE and PCPs into a similar structure without introducing clinically significant changes.
Specifically, differential diagnoses were truncated to the same length for each patient case as AMIE tended to produce longer differentials on average which may have provided a hint to clinical evaluators as to the provenance of the differential.
Management plans from both AMIE and PCPs were automatically reformatted into a pre-defined template consisting of diagnostic steps, treatment steps, and follow-up steps using an LLM-based transformation step with Gemini 2.5 Pro.
It was this truncated differential diagnosis and reformatted management plan which was assessed for AMIE and PCPs respectively in this rating step.
A manual audit of 270 outputs by the study team ensured semantic integrity after these transformation steps before clinical evaluators conducted the rating in a blinded and randomized presentation (\cref{appendix:blinding}).
Clinical evaluators rated the quality of management plans and differential diagnoses from AMIE and PCP both in a comparative manner, and a standalone pointwise manner.
To assess the effectiveness of the blinding procedure, raters were asked to make a guess as to the provenance of each output.
Rating scales are provided in \cref{appendix:tab:clinical_evaluator_rubric_diagnosis_and_management}.
For each patient case, ratings were provided by a panel of three clinical evaluators separately and results are based on the median rating across the three clinical evaluators.

\subsubsection{Exploratory outcomes}

We sought to explore whether the measurement of AMIE's diagnostic accuracy varied according to the extent of actual post-consultation investigations that the PCP required to establish the eventual definitive diagnosis. For example, after the PCP visit following the AMIE consultation, in some cases the PCP was able to reach a diagnosis for a self-limiting condition without the patient requiring a further investigation or appointment. However, in other cases, the PCP's final diagnosis required a series of investigations and further consultations; this could increase the discrepancy in information needed to establish the definitive diagnosis compared to what was available to AMIE at the pre-encounter patient-AMIE interaction.

After completion of the study and extraction of final diagnoses, a board-certified internist (AR) reviewed all patient charts and labeled the diagnostic method by which final diagnosis was determined---presumptive (made by PCP without further testing), specialist (made on referral to a subspecialist), or made via diagnostic testing (i.e., laboratory, microbiological, pathological, or imaging).
Multiple labels could be applied, for example, a specialist may use a diagnostic test to obtain the correct diagnosis.
A limitation of our methodology is that urgent care complaints can be self limited, meaning they resolve with time without a specific medical intervention, and may not require further testing or follow-up, especially if symptoms self resolve.
In these cases, the final diagnosis corresponded to a presumptive diagnosis as formulated by the PCP, presenting a less robust reference standard compared to final diagnoses that were corroborated by diagnostic testing and/or specialist follow-up.
We performed subgroup analyses for AMIE's diagnostic accuracy based on the type of the final diagnosis to understand potential differences.

Because AMIE produces thinking traces and intermediate differentials between conversational turns, it was possible to examine the internal state of the model over the course of the conversation.
To facilitate this analysis, we used Gemini 2.5 Pro in both an extractive and evaluative manner.
For each turn of the conversation, the differential and predicted probability for each diagnosis were extracted.
With the final diagnoses in context, a Gemini 2.5 Pro-based auto-rater was used to assess the correctness of each item on the differential at each turn as well as the Bond/Graber rating \cite{bond2012differential} for overall differential diagnosis quality.
We used these results to explore various aspects of the model's internal diagnostic reasoning over the course of the conversation, including its confidence level and diagnostic accuracy over time.

\subsection{Data Analysis}

Inclusion criteria for data analysis was both (1) completion of a patient-AMIE interaction and (2) successful follow up to the scheduled urgent care appointment, to enable comparison and contextualization using information from the appointment.
Incomplete survey data did not exclude participants other than omission of their missing data from respective analyses.

For each patient case, triplicate clinical evaluator ratings were aggregated using the median.
Pointwise ratings from clinical evaluators for the quality of management plans and differential diagnoses from AMIE and PCPs respectively were compared using two-way Wilcoxon signed-rank tests with patient-level pairing, followed by Bonferroni correction across the five rating questions (differential diagnosis quality, as well as appropriateness, cost effectiveness, practicality and safety of the management plan).

For all proportions, error bars were computed using 95\% confidence intervals for binomial proportions.
This included comparative ratings from clinical evaluators, blinding outcomes, as well as diagnostic accuracy of AMIE.

Patient responses on the General Attitudes Towards AI Scale (GAAIS) were first aggregated using the mean Likert score across all individual scale items.
This was done for pre-AI, post-AI, and post-provider surveys separately.
We used Friedman omnibus tests followed by pairwise two-sided Wilcoxon post-hoc tests, with patient-level pairing, to compare the distributions of mean GAAIS scores between pre-AI, post-AI, and post-provider surveys.
Cases where any of the three patient-facing surveys were not completed were excluded from these these analyses.
This was done for the overall scale, as well as the two GAAIS sub-scales corresponding to perceived concerns and utility respectively.

For the semi-structured interviews, we analyzed session notes from interviews with patients, AI supervisors, and PCPs and then used reflexive thematic analysis to identify overarching themes from the qualitative data \cite{Braun2019}.

\section{Results}
\label{sec:results}

\subsection{Patient and PCP Participants}

A CONSORT diagram is provided in \cref{tab:participation}.A visualizing the flow of patient participants through the study.
During the study period, 1,452 urgent care visits were scheduled at HCA.
A total of 140 eligible participants provided consent verbally via phone to a member of the research team, and were sent an online consent form.
Of these, 20 did not complete the consent form, and 6 completed the form, but did not show up to the scheduled AMIE encounter, resulting in 114 patients who signed the consent form and initiated the AMIE encounter.
Of these, seven participants exhibited protocol deviations or patient technical issues, five participants were not able to complete the study due to system failures, and two participants met exclusion criteria during the chatbot conversation (one pregnancy related, one mental health related), resulting in an AMIE interaction completion rate of 87.7\% (100 of 114). In the two cases of exclusion criteria, the chief complaint documented in the chart for the reason for the urgent care visit was discordant with what the patient truly wanted to discuss. 
As pre-specified, enrollment concluded once 100 patients had completed an interaction with AMIE.
Two of these patients did not show to their scheduled urgent care PCP appointment. One patient was not located in Massachusetts for her scheduled telehealth urgent care appointment (a Massachusetts law) and thus their appointment was canceled. The second patient had symptoms that resolved by the time of the urgent care appointment. Chart review revealed that both patients had later followed up with their PCP and experienced no harm as a result of missing the scheduled appointment. 

The group of consented PCPs participating as providers in this study included a total of 11 attending physicians, 61 resident physicians, and 5 nurse practitioners. 
A subset of 5 attending physicians were involved in the initial set of 10 patients seen during the pilot phase.
Not every consenting PCP had a patient recruited into the the study. There were 62 patients scheduled to be seen by attendings, 26 by residents, and 12 by nurse practitioners.

The 98 patients who had completed both the AMIE encounter and the PCP encounter were included for data analysis.
Patient survey completion was high with approximately 90\% response rates across the pre-AMIE (91.8\%), post-AMIE (90.8\%), and post–urgent care appointment surveys (89.8\%).
AI supervisors completed 100\% of surveys.
PCPs had the lowest survey completion rate at 61.2\%.
Details on survey completion rates are provided in \cref{appendix:survey_details}.
Of the 60 surveys completed by PCPs, 16 (27\%) included the selection they did not have a chance to review the AMIE transcript or summary prior to the urgent care appointment.

\clearpage
Patient demographic data were collected voluntarily via pre-AMIE surveys (\cref{tab:participation}.b).
Eight of 98 patients did not complete the pre-AMIE survey.
We include these eight patients in counts for `Prefer not to say' and `N/A' responses.
Across the 98 patient participants, 51\% were below the age of 50, 39\% were 50 years or older, and 10\% preferred not to state their age range, including the eight patients who did not complete the survey.
All patients, including those who did not reveal their age range in the pre-AMIE survey were confirmed to be 18 years or above during the initial EHR-based eligibility check.
The majority of patients self-identified as women (68\%), and reported English as the language spoken at home (86\%).
The patient sample included different racial/ethnic groups, including White (49\%), Black or African American (26\%), Hispanic or Latino (8\%), and Asian (5\%).
Compared to the total 1,452 urgent care visits during the study period, patients participating in this study tended to skew towards younger ages as over half of total urgent care visits at the clinic during the study period were over the age of 60.
However, during the study period, the total urgent care visit population trended towards female (74\%) and White (52\%) populations which was consistent with the patient sample in this study.
In the study, just over half of patients (55\%) self-reported high health literacy, indicating that they felt extremely confident in filling out medical forms independently.
Slightly less than half of patients (46\%) self-reported the highest possible scores on the tech literacy scale.
A bimodal distribution was seen in the frequency of chatbot use with 38\% of participants using a chatbot less than once a month or never compared to 37\% using a chatbot multiple times per week.
Almost a quarter (24\%) of participants reported having used a chatbot for their own health prior to participating in the study.

\begin{figure}[t!]
    \centering
    \includegraphics[width=\textwidth,keepaspectratio]{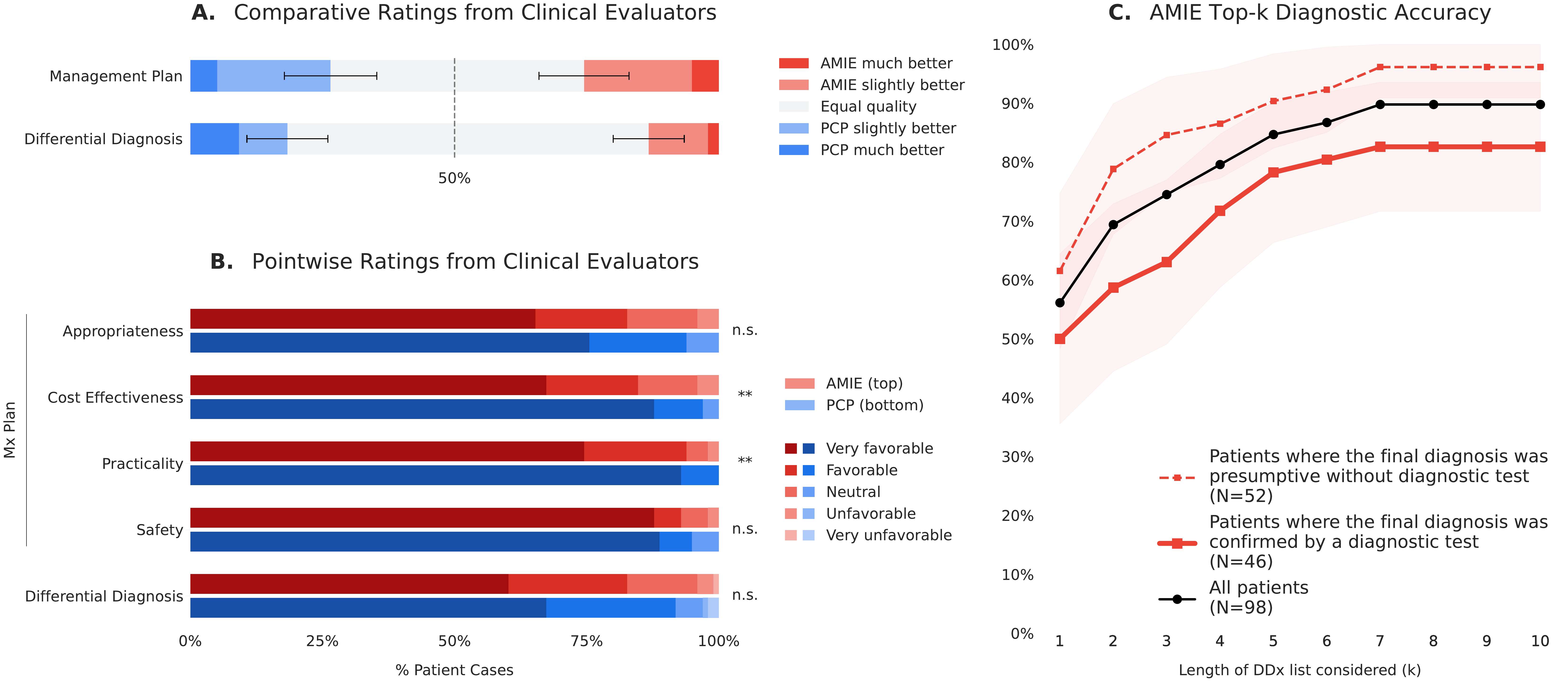}
    \vspace{0.5em}
    \caption{\textbf{Clinical Reasoning Performance.} Clinical evaluators rated the quality of management plans and differential diagnoses of both AMIE and PCPs in a blinded randomized manner. For each patient case, ratings were aggregated as the median rating across a panel of three independent clinical evaluators. \textbf{(A) Comparative ratings from clinical evaluators} assessed the quality of two candidate management plans and differential diagnoses in a side-by-side manner relative to each other; error bars for comparative ratings represent 95\% confidence intervals for binomial proportions (N=98) for expressing a preference (`slightly better' or `much better') for either of the two candidates. \textbf{(B) Pointwise ratings from clinical evaluators} assessed the quality of each management plan and differential diagnosis individually on a 5-point Likert scale. For pointwise ratings, asterisks represent statistical significance per two-sided Wilcoxon signed-rank tests with Bonferroni correction ($**:p<0.01$, $n.s.: $ not significant).
    In addition to ratings from clinical evaluators, we measured \textbf{(C) AMIE's Top-k diagnostic accuracy} as compared to the final diagnosis extracted for each patient via chart review eight weeks after their PCP visit. In addition to overall accuracy across all patients (N=98), we provide accuracy for the subset of patients where the final diagnosis was confirmed by a diagnostic test such as imaging, microbiology, laboratory, pathology, EKG (N=46), and the subset where this was not the case, i.e., where the diagnosis was presumptive without a diagnostic test, irrespective of whether the diagnosis was made by a PCP or specialist  (N=52). Error bars for diagnostic accuracy correspond to 95\% confidence intervals for binomial proportions.}
    \label{fig:clinical_reasoning_performance}
\end{figure}

\subsection{Safety}

Across all patient-AMIE interactions in this study, zero safety stops were required by the group of AI supervisors overseeing these interactions, per the four pre-specified safety criteria:
(1) immediate concern for harm to self or others, (2) significant emotional distress exhibited by the patient related to the AI interaction, (3) potential for clinical harm identified by the supervisor based on the conversation, or (4) an explicit request from the patient to end the session. On three occasions, the AI supervisor made remarks to the patient during or at the conclusion of the patient-AMIE interaction. One case was to clarify symptoms to rule out a potentially emergent condition which the patient did not have, one case was to clarify contingency criteria to seek emergency care, and one case was a correction of AMIE stating that a date a patient had surgery was in the future when it in fact was in the past as of the date of the patient-AMIE encounter. 

\subsection{Clinical Reasoning Performance}

\cref{fig:clinical_reasoning_performance} provides an overview of clinical reasoning performance as assessed by clinical evaluators rating the quality of management plan and differential diagnoses from both AMIE and PCPs, as well as AMIE's top-k diagnostic accuracy with respect to the final diagnosis for each patient case.

Comparative ratings from clinical evaluators (\cref{fig:clinical_reasoning_performance}.a) suggest AMIE and PCPs had similar overall quality of their management plans and differential diagnoses.
This result was also supported by pointwise ratings from the same clinical evaluators (\cref{fig:clinical_reasoning_performance}.b).
For pointwise ratings, a statistically significant difference was not observed between AMIE and PCPs for the appropriateness (p = 0.1) and safety (p = 1.0) of their respective management plan, as well as the quality of their differential diagnoses (p = 0.6).
However, management plans from PCPs were rated significantly more favorably for cost effectiveness (p = 0.004) and practicality (p = 0.003), compared to AMIE's management plans.
The rate of clinical evaluators making a guess as to the provenance of diagnoses and management plans (AMIE or PCP), and guessing correctly was 59.18\% (95\% CI: 49.45\%, 68.91\%; N=98).

AMIE's top-k diagnostic accuracy with respect to the final diagnosis extracted from chart review is shown in \cref{fig:clinical_reasoning_performance}.b.
Based on Bond/Graber ratings $\geq 4$, AMIE included the final diagnosis in 88 of 98 cases (90\%) within the first 7 candidates of its ranked differential, and in 73 of 98 cases (75\%) within the first 3 candidates.
AMIE's top diagnostic candidate matched the final diagnosis in 55 of 98 cases (56\%).
AMIE's diagnostic accuracy for all values of possible lengths of the differential are provided in \cref{appendix:tab:ai_top_k_ddx_accuracy}, and the distribution of Bond/Graber scores is provided in \cref{appendix:tab:bond_graber_scores}. The comparison of diagnoses seen in the study and diagnoses in the overall urgent care population during the study period is shown in \cref{appendix:diagnostic_demographics}.
AMIE's diagnostic accuracy remained at a high level for the subset of patients where the final diagnosis was confirmed by a diagnostic test (N=46), though trending higher for the other subset of patients where the final diagnosis was presumptive without diagnostic testing involved (N=52).
In \cref{appendix:turn_analysis}, we provide a turn-level analysis of AMIE’s working differentials. We demonstrate that AMIE generates accurate differentials early in the interaction and exhibits a similar reduction in diagnostic uncertainty over time, regardless of whether its final differential is correct.

\begin{figure}[t!]
    \centering
    \includegraphics[width=0.8\textwidth,keepaspectratio]{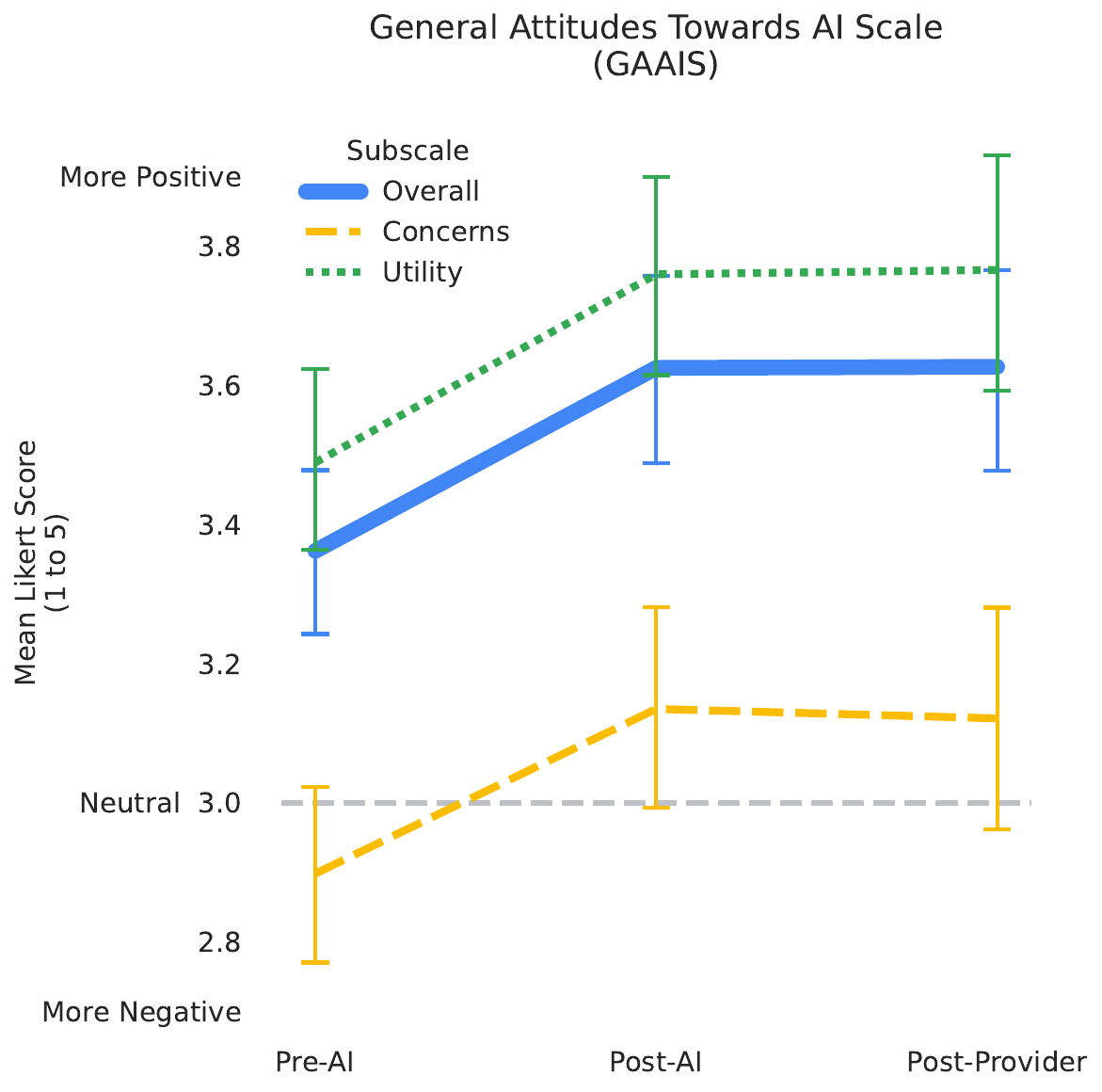}
    \vspace{0.5em}
    \caption{\textbf{Effect on patient attitudes towards AI.} Patients completed the General Attitudes towards AI Scale (GAAIS) prior to interacting with AMIE (Pre-AI), after interacting with AMIE (Post-AI), and after the urgent care consultation with the provider (Post-Provider). The GAAIS scale includes two sub-scales corresponding to (1) perceived utility and (2) concerns around AI. Attitudes shifted more positive after interacting with AMIE and remained at an elevated level after seeing the PCP. This change was statistically significant for both sub-scales and the overall scale as confirmed by a Friedman omnibus test followed by pairwise Wilcoxon post-hoc tests (p < 0.001 for omnibus and pairwise tests).}
    \label{fig:gaais}
\end{figure}

\subsection{Patient Experiences}

\textbf{Surveys.} Patient attitudes towards AI (\cref{fig:gaais}), as measured by mean GAAIS scores, shifted more positive after interacting with AMIE and remained at an elevated level after seeing the PCP.
This change was statistically significant as confirmed by a Friedman omnibus test (p < 0.001 for overall scale and both sub-scales) followed by pairwise Wilcoxon post-hoc tests.
The same finding applied to the overall scale (Pre-AI vs. Post-AI: p < 0.001; Post-AI vs. Post-Provider: p = 0.86), as well as the two sub-scales, corresponding to perceived concerns (Pre-AI vs. Post-AI: p < 0.001; Post-AI vs. Post-Provider: p = 0.93) and perceived utility (Pre-AI vs. Post-AI: p < 0.001; Post-AI vs. Post-Provider: p = 0.90) respectively.
Patient survey responses for GMCPQ, PACES and PCCBP rubrics are visualized in \cref{fig:amie_conversation_quality} and reported in \cref{{sec:amie_conversation_quality}}.

\textbf{Interviews.} From the larger study cohort (N=100), 20 participants took part in remote, moderated UX follow-up interviews to evaluate the AMIE interaction. This subset consisted of a predominantly female (80\%) and White (55\%) population, with minimal representation from Black or African American (15\%) and Asian (10\%) communities. Using reflexive thematic analysis  \cite{Braun2006-nt}, we identified several themes describing how participants experienced and interpreted their interactions with AMIE. Participants were primarily motivated to use the chatbot by the perceived novelty and utility of conversational AI, such as to explore new technology and to better prepare for time-limited clinical encounters by organizing their thoughts beforehand. They viewed AMIE as a means to organize their health narrative and provide additional context before a PCP visit. Additionally, patients appreciated the system's detailed history-taking, ensuring that providers receive a specific description of patient complaints rather than a checklist or questionnaire typically administered through patient portals or clinic staff prior to visits. Initially, participants were skeptical of the chatbot’s capabilities, but as they continued their conversation with AMIE, many patients praised the overall performance. Across interviews, participants frequently described AMIE as empathetic and human-like, sharing it did a good job of detailed history-taking and using plain and easy-to-understand language which made the participants feel understood and validated. Comparisons with human healthcare providers highlighted AMIE’s accessibility and approachability, particularly its use of plain language and its ability to deliver health information without the intimidation or pressure often felt in clinical settings. Participants also reported that AMIE positively impacted subsequent PCP visits by facilitating communication, such as providing a patient conversation history, and enhancing trust. While patients appreciated the benefits of using conversational AI, they expressed concerns about data privacy and AMIE's ability to manage complex or high-risk health issues. Participants also emphasized clear boundaries for appropriate uses of the chatbot, such as for primary care inquiries rather than emergencies. Users identified opportunities for improvement, such as clearer data transparency and safety guardrails for future iterations. 

\subsection{{Clinician Experiences}}

\subsubsection{Providers}

\textbf{Surveys.} Provider post-survey responses are provided in \cref{tab:provider_post_survey_responses}. Of the 60 surveys completed by PCPs, 16 included the selection that the PCP did not have a chance to review the AMIE transcript or summary prior to the urgent care appointment. Among the remaining 44 surveys, PCPs indicated that they found preparing the urgent care appointment with AMIE helpful in 75\% of cases (41\% `very' and 34\% `somewhat' helpful) and harmless in 68\% of cases (57\% `harmless' and 11\% `somewhat harmless'), with neutral responses in 16\% of cases for helpfulness, and 30\% for harmfulness. PCPs assessed the preparation with AMIE as somewhat unhelpful in only four cases and somewhat harmful in a single case, with zero selections of very unhelpful or very harmful across all completed surveys. PCPs indicated that they trust the information from the AMIE conversation in 64\% of cases (23\% `strongly agree' and 41\% `agree'), with neutral responses for 32\% of cases, and only a single selection where the PCP disagreed with that statement and none who disagreed strongly. Regarding the question of whether preceding the urgent care appointment with the AMIE-chat may have changed the PCP's behavior or actions during the urgent care consultation, PCPs thought that this was the case in 57\% of cases (21\% `definitely' and 36\% `probably' yes) and that it was not the case in 18\% of cases (7\% `definitely' and 11\% `probably' no), while PCPs were unable to decide or made no selection in 25\% of cases.

\clearpage
\textbf{Interviews.} The ten PCPs with the highest volume of patients in this study participated in semi-structured interviews.
Interviews with PCPs sought to probe the clinical utility and workflow impact of the AMIE transcript and summarization. The qualitative data revealed a consistent pattern: PCPs view AMIE as a tool that shifted the visit dynamic from data gathering to data verification and expanded opportunities to further explore health concerns. PCPs likened AMIE’s transcripts and summaries to that of a third-year medical student, noting that the pre-visit summary often exceeded the detail of standard intake notes. 
Furthermore, interview data highlighted a positive impact on the patient-provider interaction.
PCPs observed that patients who interacted with AMIE prior to the visit arrived prepared and organized, with their thoughts and questions presented in a coherent narrative.
Unlike experiences with patients using other online tools to search their symptoms, in which patients were described by PCP as bringing dissonant pieces of information not tailored to them, patients who interacted with AMIE were described as having a more personalized understanding of their health concerns, which reduced anxiety and resulted in well-formed stories to describe their concerns to the provider.
This effect on patient preparedness enabled the healthcare visit to shift from information-gathering interviews to more collaborative conversations, thereby improving shared decision-making.
PCPs reported being able to build rapport more quickly and to focus more of the visit on counseling and management.
Overall, PCPs agreed that AMIE is a valuable tool that enhances their clinical practice by streamlining the care process and allowing them to maintain and nurture their rapport with their patients.

\subsubsection{AI Supervisors}

All seven safety supervisors were invited to interviews after the study, and five agreed to participate in interviews, to share their experiences overseeing the patient-AMIE encounter and opinions on the real-world deployment of AMIE.
Supervisors described strong clinical and conversational performance once the chatbot interaction commenced, following initial logistical challenges.
Supervisors identified system setup and using study-specific username and password combinations required to login as key barriers, including microphone and screen sharing permissions, and varying levels of patient technical literacy, requiring real-time guidance for some patients during the initial setup.
Despite these barriers, supervisors reported that patients engaged rapidly once connected.
Interactions were described as conversational, intuitive, and largely self-sustaining, allowing supervisors to adopt a predominantly observational role.
Clinically, AMIE was regarded as thorough and generally aligned with supervisors’ interpretations of the case.
Noted strengths included structured history-taking, appropriate follow-up questions, and the ability to redirect unfocused narratives toward key clinical details.
Supervisors noted that older patients often required long interaction times due to both logistical barriers (i.e., slow typing) and clinical considerations (i.e., long medication lists, complex medical history). Hypothetical issues related to safety and governance were also highlighted, including triage accuracy in urgent cases, the risk of over-validating health anxiety, and broad differential lists in complex cases.
Supervisors recognized potential efficiency gains in pre-visit history collection and summarized outputs, but cautioned against overreliance by patients or clinicians and diagnostic anchoring.
Overall, supervisors characterized AMIE as a promising, resident-level tool that performs optimally with clear supervision, improved onboarding, and enhanced safeguards for high-risk scenarios.

\begin{figure}[ht!]
    \centering
    \includegraphics[width=\textwidth,keepaspectratio]{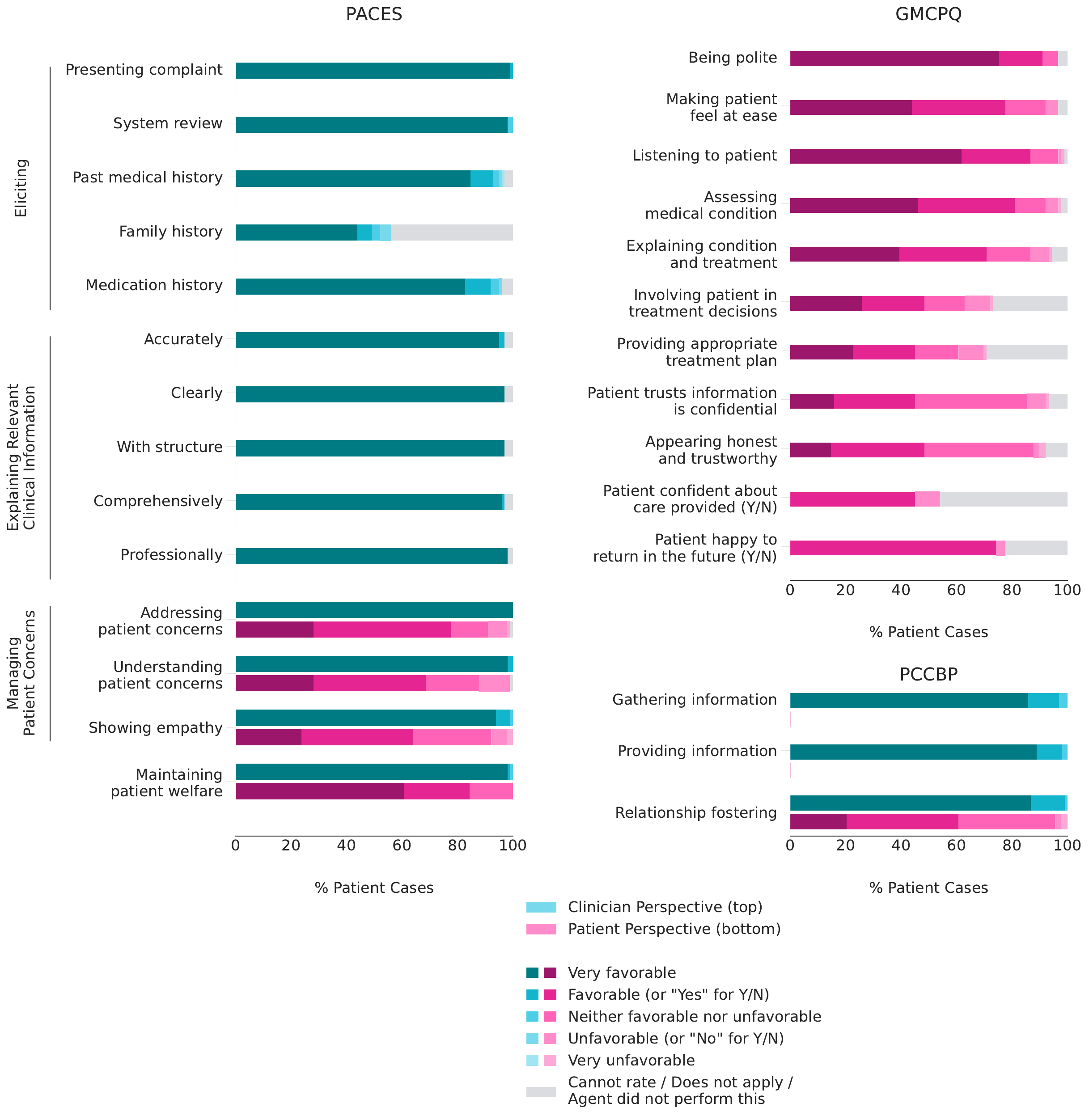}
    \vspace{0.5em}
    \caption{\textbf{AMIE Conversation Quality.} The quality of AMIE conversations was rated from patient and clinician perspectives. Patient perspectives were collected through surveys immediately after patients completed their interaction with AMIE and the safety debrief with the AI supervisor. Clinician perspectives were rated post-hoc by a panel of three independent clinical evaluators per patient case, and aggregated using the median across the three ratings per case.}
    \label{fig:amie_conversation_quality}
\end{figure}

\clearpage
\subsection{AMIE Conversation Quality}
\label{sec:amie_conversation_quality}

Ratings of AMIE's conversation quality from clinician and patient perspectives are visualized in \cref{fig:amie_conversation_quality}.

\textbf{Clinician Perspective.} Clinical evaluators rated the quality of AMIE's clinical dialogue using the PACES and PCCBP rubrics.
Overall, AMIE's consultations were rated as favorable or very favorable for the majority of patient cases across all criteria, with no more than three cases of 98 total with borderline ratings for any criteria.
Unfavorable ratings were uncommon and were limited to inadequate past medical history for two cases, family history in four cases, medication history in a single case.
In almost half of cases (43 of 98), clinical evaluators determined that omission of family history was acceptable.
Detailed counts for clinical evaluator ratings are tabulated in \cref{appendix:tab:clinical_evaluator_ratings}.

\textbf{Patient Perspective.} Patients assessed AMIE's conversation quality using subsets of PACES and PCCBP rubrics, as well as the GMCPQ rubric.
For PACES and PCCBP criteria assessed by both clinical evaluators and patients, clinical evaluator ratings trended more positive compared to patient ratings.
Despite this discrepancy, patients also rated AMIE's conversation quality as favorable or very favorable for the majority of patient cases across all PACES and PCCBP criteria.
The same trend of majority favorable or very favorable assessments was observed for all but two GMCPQ criteria, excluding cases marked by patients as `Cannot rate / Does not apply / Agent did not perform this'. The two criteria where patients assigned larger proportions of ratings indicating neither favorable nor unfavorable were trusting information confidentiality and perceived honesty and trustworthiness. 
Across all GMCPQ criteria, unfavorable ratings made up the smallest proportion at <10\% of cases.
Across four GMCPQ criteria representing aspects less applicable to the specific study setting (involving the patient in treatment decisions, providing an appropriate treatment plan, patient being confident about the care provided, willingness to return in the future), larger proportions of patients (over 20\%) selected  `Cannot rate / Does not apply / Agent did not perform this'.

\section{Related Work}
\label{sec:related-work}

\subsection{Differential generators}

Efforts to develop clinical decision support systems have spanned decades, long predating modern machine learning or generative AI systems \cite{Miller1994}.
The Cornell Medical Index (CMI), developed in 1949 \cite{Brodman1949}, captured information not elicited during physicians’ history-taking that proved pertinent in the diagnostic reasoning process. \citet{Ledley1959} introduced the principles of leveraging logic, probability, and value theory in medical reasoning to make appropriate diagnostic and management decisions.
This conceptual foundation laid the groundwork for subsequent systems like INTERNIST-I, a computerized diagnostic tool developed in the 1970s that could make complex diagnoses in internal medicine, and its successor Quick Medical Reference (QMR) \cite{Miller1982, Miller1986}.
Diagnostic decision support systems, commonly termed differential generators, such as DxPlain and Isabel performed comparably with expert consensus, though with limited real-world impact on patient care \cite{Feldman1991, Barnett1987, Martinez-Franco2018, Riches2016, Hautz2025}.
Early evaluations of LLMs suggested performance consistent with differential generators \cite{Feldman2025, Bridges2024, kanjee2023accuracy, McDuff2025}, and modern reasoning models have far outpaced these systems \cite{brodeur2025superhumanperformancelargelanguage}.

\subsection{Patient-facing artificial intelligence}

Patient-facing AI has evolved tremendously over the last decade.
Prior to the advent of LLMs, Babylon Health’s Triage and Diagnostic System represented a pioneer in this realm, offering an AI-powered symptom checker that could query and triage patients’ symptoms to provide appropriate next steps \cite{Baker2020}.
However, evaluation revealed varying diagnostic accuracy, particularly within specialty care \cite{Baker2020, Gilbert2020, Gehlen2025}.
LLMs have drastically advanced the landscape of patient-facing AI, now offering chatbots that can engage in realistic human conversation, gather comprehensive histories, and generate nuanced differentials \cite{Tu2025-lf,palepu2025conversationalaidiseasemanagement}.
Early data from these implementations---largely in telehealth settings---have been promising \cite{Gilbert2020, Kopka2025, Mukherjee2024}.
Patient communication with AI integrated into health advice lines was shown to be feasible and generally safe \cite{Lize2025}, and a recent retrospective evaluation by \citet{Zeltzer2025-qi} showed that an AI system built from an ensemble of discriminative machine-learning models and augmented with rule-based logic, produced diagnostic and management plans that could potentially meaningfully contribute to patient interactions as an intake tool for telehealth urgent care complaints.
LLMs are similarly helpful for care navigation; patients interacting with LLMs for history taking prior to specialist consultation improved efficiency of consultations and perceived care coordination \cite{Tao2026-nl}.   

\subsection{Oversight systems for patient-facing AI}

As patient-facing AI systems grow increasingly capable, appropriate oversight systems may be needed to ensure safe and appropriate research deployments \cite{Weissman2025}. This call for safety is highlighted by recent work suggesting chatbots that are disconnected from the healthcare system without human oversight may undertriage patients and unreliably address mental health concerns \cite{Ramaswamy2026}. 
Some have proposed frameworks for AI systems that mirror existing supervisory models for advanced practice practitioners \cite{Morrell2022}.
Earlier Bayesian models, such as Babylon Triage and Diagnostic System  \cite{Meyer2020, Baker2020} operated autonomously, sparking concerns over potential misdiagnoses and patient harm \cite{Fraser2018}.
LLM-based chatbots have largely adopted a human-in-the-loop supervisory system, where transcripts are reviewed by licensed clinicians and prescriptions are provided only under physician review \cite{Zeltzer2025-qi}.
Hippocratic AI's agents are designed to hand off to nurses in a human-in-the-loop system, but they only engage in non-diagnostic, low risk tasks \cite{Mukherjee2024}.
Other models have deployed stricter oversight, including Alan Health’s chatbot Mo, which demands a rating of each of the chatbot’s replies by a physician within 15 minutes \cite{Lize2025}, and Therabot, where all chatbot responses are monitored and patients are contacted in the setting of inappropriate responses or immediate safety concerns \cite{Heinz2025}.
Prior analyses of AMIE in a simulated setting evaluated decoupling information gathering from provision of medical guidance, concluding each clinical encounter after obtaining a sufficient history and providing a differential and management plan only to the overseeing physician.
Using this framework, AMIE outperformed nurse practitioners, physician assistants, and PCPs in providing valuable intake, differential diagnoses and management plans to an overseeing PCP \cite{Vedadi2025}.

\subsection{Patient perspectives in patient-facing AI}

Beyond safety and efficacy, patient trust in patient-facing AI systems is critical for successful deployment.
Prior work has demonstrated varying levels of acceptability of AI systems among patients, with greater patient comfort in applications demanding less (e.g., digital scribe) compared to more (e.g., virtual avatar) patient interaction \cite{Foresman2025}.
In a study by \citet{Moore2025}, interviews with patients who had used a chatbot integrated within a health system’s electronic health record reflected a consensus that the tool was generally helpful.
Moreover, some participants reported increased trust due to its convenience and lack of judgment, particularly among demographic groups historically distrustful of the health care system \cite{Moore2025}.
In a similar study, patients were especially drawn to the chatbot for administrative or sensitive tasks \cite{Dellavalle2025}.
A scoping review on patient perspectives reflected shared impressions that AI-powered systems could improve diagnostic accuracy, reduce human bias and error, and improve care access;
conversely, AI’s reliability, loss of personal human connection, and lack of transparency and patient autonomy remained important concerns for its use \cite{Osnat2025}. 

\section{Discussion}
\label{sec:discussion}

In this work, we performed a feasibility and safety study of a patient-facing conversational AI engaging in urgent care visits in a real world ambulatory primary care setting.
We found the deployment of AMIE in a real patient care workflow to be practical with a high patient-AMIE interaction completion rate and follow-up to scheduled urgent care appointments.
Additionally, under the supervision of human physicians, we found that conversations with AMIE did not produce any safety alerts based on prespecified study criteria.
The quality of AMIE's communication with patients was highly rated among clinical evaluators along with positive patient sentiment towards AMIE and conversational AI involvement in their care.
In this real world setting, we demonstrated PCPs and AMIE had overall similar quality of their differential diagnoses and management plans, with PCPs receiving better ratings on aspects of cost effectiveness and practicality.

\subsection{Real World Implementation}
This study marks the first prospective real-world evaluation of an LLM-based conversational AI agent performing a text-based urgent care visit under real-time supervision of a dedicated safety physician.
Prior studies have evaluated conversational AI with real patients but have been either retrospective \cite{Zeltzer2025-qi} or limited to advice lines, telehealth, or intake chats prior to a specialist consultation  \cite{Lize2025, Zeltzer2025-qi, Tao2026-nl}.
In contrast, this study included all-comers inclusive of both telehealth and in-office evaluations.
Conducting the study in a high-volume academic medical center makes the findings more reflective of real-world implementation within busy clinical workflows.
During the study period, approximately 10\% of all urgent care visits were enrolled in the study, with patient demographics of study participants skewing towards younger ages, though otherwise largely matching that of the overall urgent care clinic population suggesting generalizability.
Regarding the patient-AMIE interaction's effect on subsequent medical care, only two patients (2\%) did not proceed to their scheduled urgent care appointment with a PCP, though there may have been a form of self-selection bias of patients highly engaged in care. 

While we found the integration of AMIE to be practical in a busy real world clinical workflow, this was not without challenges.
The oversight setup in this study implemented live supervision by a remote physician through screen-sharing of computer screen by the patient via a secure video call.
Technical barriers related to this oversight setup were a consistent theme for both patients as well as operationally.
Roughly 7\% of enrolled patients could not complete the study due to device related issues and upon inquiry of AI supervisors, many patients required significant technology onboarding prior to commencement of the AMIE interaction.
These findings are concordant with known health equity barriers of digital health such as low technology literacy and limited access to an adequate device \cite{Takahashi2022, Creber2023}.
Our reported rate of interaction incompletion due to patient related technology barriers likely underrepresents the magnitude as our recruitment process deliberately screened out those without an adequate device (laptop or desktop computer) which likely naturally skewed towards a younger patient population.
On the operational side, PCPs were able to review the transcript ahead of the scheduled clinic visit only 73\% of the time among the cases where PCPs completed the survey.
In this study setting, AMIE was not integrated with the clinic's EHR, but was operated as a separate web application in a secure environment.
The successful transmission of transcripts to PCPs, therefore, required lead time due to several reasons such as time necessary to securely store the data on clinic infrastructure, and an email of the transcript from clinical research staff to PCPs.
With scale and stronger integration, we would anticipate this process to increase in efficiency along with automation to push AMIE conversations to PCPs in a streamlined manner.

\clearpage
Some PCPs expressed that receiving the information from an AMIE pre-consultation asynchronously from the time they usually allocated for pre-clinic preparation was inefficient, because it provoked multiple reviews of the same patient. Evidence supports that workflows which allow physicians to prepare and personalize a patient encounter results in increased visit efficiency and connection with patients \cite{Tao2026-nl, Zulman2020, Vedadi2025}, offering considerable promise that with workflow adaptation this qualitative observation could be addressed to improve physician experience.
As LLM technology improves and earns physicians’ trust, hybrid workflows may adapt in ways that allow clinicians richer pre-visit preparation, prior to meeting with patients that have also had significant helpful preparation for a fruitful encounter.

\subsection{Conversational Safety}

There were no safety stops across all 100 patient-AMIE conversations, reflecting that the system did not elicit concerns for immediate patient harm, emotional distress, potential for clinical harm, or an explicit request by the patient to stop the interaction.
These prespecified safety criteria were specifically developed to cover a broad range of potential safety concerns and purposely left flexibility for multiple scenarios to qualify as safety stops.
The safety approach deployed in this study was rigorous and conservative as all 100 patient-AMIE interactions had continuous, real-time human oversight by a physician.
Our findings are consistent with other real world evaluations of conversational AI which similarly report low safety intervention rates \cite{Lize2025, Heinz2025}.
However, unlike previous real world studies \cite{Lize2025, Zeltzer2025-qi, Tao2026-nl} our study gauged patient safety for an interaction where the AI chatbot would not only collect information from the patient, but also conclude the conversation by producing possible diagnoses and next steps for the underlying health concern framed as topics the provider may want to discuss with the patient during the visit.

The nature of safety supervision (with patient foreknowledge that they were being observed, even though the supervisor had their camera and microphone disabled) makes it possible that some of the high performance may be due to the Hawthorne effect.
For example, foreknowledge of observation likely limited adversarial prompting, such as presenting evidence they found on the internet regarding their condition, which has been known to alter the safety profile of LLM output \cite{Lee2025}.
Additionally, these safety results need to be taken within the context of screening out pregnant patients, those with mental health chief complaints, or those requiring emergency care.
Directing potentially medically unstable patients to emergency care is an important task in any triage system.
No patients included in this study were triaged to emergency care settings and thus the ability of AMIE to safely navigate this scenario in the real world remains unstudied.
However, AMIE has been trained to suggest immediate action such as an emergency room visit similar to other conversational AI agents \cite{Tu2025-lf,palepu2025conversationalaidiseasemanagement}.
Nonetheless, this study provides empirical evidence that conversational AI can be safe in patients presenting with most urgent care complaints.
These findings serve as motivation for larger scale trials with close human involvement to ensure safety as this technology begins its infancy in real world patient care workflows. 

\subsection{Dialogue Quality}

As graded by physicians, AMIE's conversational performance was rated as very favorable on several axes evaluating history gathering, explaining clinical information, managing patient concerns and relationship fostering with patients.
These ratings outpaced that of prior work in a simulated setting, albeit with an older model, suggesting that standardized patient results translate into real world settings \cite{Tu2025-lf}.
The favorability noted in this study is also in alignment with prior real world studies which also found physicians strongly approve of conversational quality from AI \cite{Lize2025}.
These ratings may partly reflect clinicians’ awareness of gaps in routine history-taking, given evidence that physicians frequently miss key history elements and often interrupt patients within seconds \cite{Ramsey1998, Marvel1999}.
The positive ratings also provide additional face validity for AMIE’s conversational quality, consistent with literature that identifies emotional intelligence, uninterrupted listening, and patient connection as central to improving clinical encounters \cite{Zulman2020}.
Overall, AMIE's strong performance as graded by physicians is a real-world confirmation of the ability of LLMs to communicate effectively, consistent with existing benchmarks \cite{Arora2025, Bedi2026}. 

Four items on the PACES rubric were graded by both blinded physicians as well as patients.
In all four domains---addressing patient concerns, understanding patient concerns, showing empathy, and maintaining patient welfare---physicians consistently rated AMIE higher than patients did.
This discordance is consistent with the larger literature on patient and physician experiences in health encounters, where there is little correlation between the views of each group \cite{Rottele2020-lm}.
These patient ratings also trended lower than those from actor patients in previous AMIE studies. 
One potential explanation for this discrepancy is that lower ratings of patient experience in this real-world study may reflect fundamental limitations in simulated (OSCE) settings for patient-facing AI systems. 

Patient ratings of conversational quality were generally high, especially in being polite, making the patient feel at ease, listening to the patient, and explaining condition and treatment.
Ratings were lower in involving patients in treatment decisions, and providing appropriate treatment plan---both tasks that AMIE was explicitly instructed not to perform but to handoff to the patient's PCP.
Patients had persistent concerns about trust---less than half felt that the system would keep information confidential or that it appeared honest and trustworthy.
However, the majority would be happy to interact with AMIE again in the future.
Overall, these findings are consistent with patient quality ratings in experimental standardized patient settings \cite{Tu2025-lf}, with the noted exception of trust, where real-world numbers are consistently worse than simulation.
These themes were echoed in our qualitative interviews, where patients suggested the importance of transparency into safety guardrails and around how data was transmitted to their PCPs in a potential setting where an AI system like AMIE might be used outside of a study context with explicit informed consent.
Trust in AI in healthcare is multidimensional; qualitative and survey research suggests that patient trust in medical AI systems flows from relationships with their human providers \cite{Yao2025-xu, Busch2025-ad}.
Experience from other telehealth interventions suggest that trust is also built over time with repeated interactions, though early experiences with technology will continue to influence the experience of clinical deployments \cite{Palakshappa2024-pr}.
We see some early evidence of this in our study---in the General Attitudes Toward AI Scale, trust increased after using AMIE and remained elevated after the patient's visit with their physician \cite{Schepman2020-aq}.
The trend held with both negative (concerns) and positive (utility) subscales, with significant and persistent elevations in attitude after just a single AMIE encounter.
Future research will not only need to explore what features and characteristics of interactions can build patient trust and confidence, but also how AI interactions can be better integrated into workflows with trusted physicians in a way that enhance the patient-physician relationship. 

\subsection{Clinical Reasoning Performance}

Prior studies of AMIE in simulated settings revealed superior diagnostic and management capabilities compared to PCPs \cite{Tu2025-lf, palepu2025conversationalaidiseasemanagement}.
However, these studies were limited in that PCPs were constrained to text-only interactions with patients which is not consistent with real-world practice  \cite{DeAlbornoz2022}.
In this study, we aimed for a `fair' head-to-head comparison by evaluating each modality in its most natural form:
a patient-LLM interaction delivered through a chatbot interface versus real-time patient interaction with a human physician, either in person or via a telehealth platform.
We found no significant difference in the overall quality of the differential diagnosis and management plan from AMIE versus PCPs.
However, scoring for the management domains of cost effectiveness and practicality favored PCPs.

The single-arm feasibility nature and safety oversight system of our study offers challenges to meaningfully evaluate the secondary outcomes of our study, comparisons of diagnostic and management quality of humans and AI.
Real-world diagnosis and management requires embodied cognition \cite{Daniel2020-qy} as well as significant chart review directly from the EHR \cite{Crawford2019-av}.
While our study independently evaluated AMIE and PCPs within their context of care delivery, our head-to-head evaluation method favored physicians who had more context, including the AMIE chat transcript and summary.
AMIE did not have access to the patient's EHR, did not have the ability to perform a physical exam, or integrate multimodal user input such as the tone of a patient's voice.
The physical exam is a core component of the diagnostic process often driven by hypotheses from history taking \cite{Garibaldi2018}.
Compared to PCPs, AMIE's differential diagnoses were generally longer prior to truncation for blinding purposes.
The limited context available to AMIE likely contributed to AMIE correctly assuming a broader differential diagnosis with subsequent inferiority in cost-effectiveness and practicality of workup compared to PCPs.
In contrast, PCPs may have been able to use this context advantage to construct a more precise differential and workup.
The results suggest that AMIE requires further alignment to aspects of cost-effectiveness and practicality of care decisions which may differ between contexts of care.
However, LLMs are already capable of  integrating and reasoning over user multimodal input \cite{saab2025advancingconversationaldiagnosticai}, including potentially live video, and early work has already shown that agents can meaningfully extract information from EHRs \cite{zakka2024almanaccopilotautonomouselectronic}.
Taken as such, the findings of AMIE's diagnostic and management capabilities in this study likely represent performance that can be enhanced through the provision of additional context and modalities, albeit with the concomitant requirement for deeper integration into EHR systems and more complex user input capabilities.
As progress continues in AI and models are more elegantly able to integrate multimodal data, are embedded into EHR environments, and develop longitudinal knowledge representations of patients, AI may be able to play an increasingly prominent role as a teammate in the care of patients beyond providing pre-visit intake. 

Recent studies have also called attention to the susceptibility of models to provide management plans that are potentially harmful to patients \cite{wu2025firstnoharmclinicallysafe}.
We found that the safety and appropriateness of AMIE management plans compared to PCPs was similar.
If models are to reach autonomous clinical workflows, safety will be paramount as prior studies have shown that patients cannot be assumed to play any oversight role \cite{Shekar2025-bt}.
It is likely that future studies with continued human physician oversight will help inform which clinical cases are safe and most suitable for AI to manage autonomously with a human-on-the-loop rather than in-the-loop. 

In order to ensure rater blinding, differential diagnoses in each case from AMIE and PCPs were truncated to match the length of the shorter differential between the two. Since AMIE tended to produce longer lists, its differential was most frequently truncated. Management plans were similarly converted into similar formats. In the evaluation of blinding methods, clinical evaluators were able to make a guess on which output was from AMIE versus PCPs in 84\% of cases, and the rate of making a guess and guessing correctly was only 59\%.
As AI is introduced into clinical practice, rigorous validation against the prevailing standard of care, a human performance baseline, should be prioritized \cite{Rodman2025, brodeur2025superhumanperformancelargelanguage}.
Robust blinding will be essential to ensure unbiased assessment, and the blinding techniques employed in this study can serve as a model for future clinical trials. 

\subsection{Exploratory Analyses}

Subgroup analyses stratified by the provenance of the final diagnosis (presumptive, specialist, or diagnostic-test confirmed) did not show significant differences in diagnostic accuracy.
In our analysis of model reasoning traces, neither model confidence nor entropy were associated with improved diagnostic accuracy---even though model confidence increased over time.
This is concordant with studies of physician cognition, where physician confidence is unrelated to both diagnostic accuracy and case difficulty \cite{Meyer2013-jq, Friedman2005-kc}.
There are additional similarities as well between model metacognition and experimental studies of physician reasoning.
Successful conversations had much higher diagnostic accuracy as early as the first conversational turn, which is concordant with findings that expert clinicians have much earlier hypothesis generation than novices \cite{Norman2017-hz}.
Successful conversations also improved with additional turns whereas unsuccessful conversations did not; the exact phenomena has been demonstrated in physicians, and likely reflects the common metacognitive bias of premature closure \cite{Krupat2017-ek}.
The most immediate implication of these findings is that merely providing reasoning traces of patient-facing LLMs to physicians is unlikely to improve their diagnostic performance, which runs contrary to arguments that the provision of chains of thought increases interpretability and trust in decision support \cite{Ayoub2026-ci}.
We intentionally did not make reasoning traces available to PCPs in our study, largely because of their length and the lack of feasibility to review prior to seeing patients.

\subsection{Future Directions}
    
This study establishes that clinical communication with conversational diagnostic AI prior to urgent care visits within a high-volume primary care practice at an academic medical center is safe and feasible.
Our results suggest several possible paths for future clinical integration.
Ambient AI scribes, for example, could allow pre-encounter AI systems to have access to and reason over vocalized physical exam or voice intonation.
Further advances in AI technology could also allow for more capable multimodal models that are able to assess the physical exam directly through a device camera within the limits of a telehealth exam.
Finally, these results suggest a potential future where in certain clinical scenarios AI agents may have a role in autonomous assessment and decision making;
however, additional studies, with focuses upon safety and identification of appropriate scenarios in real-world situations without pre-triage would be necessary, with robust safety mechanisms in place. 

\subsection{Strengths and Limitations}

Studying real-world patients in active clinical workflows is a key strength and provides early insight into operational barriers that may affect scalability in similar high-throughput settings.
The study design was also prospective: we reviewed eight weeks of clinical data to determine the most likely final diagnosis allowing rigorous evaluation of differential diagnoses;
such an approach has not been previously performed in any study evaluating conversational AI in real-world settings. We employed a robust human oversight system using physicians to continuously monitor patient-AMIE conversations for safety concerns especially given that AMIE presented potential diagnoses directly to patients.
Although this is resource intensive, it can serve as a gold standard for safety oversight of LLM output presented to patients.
Head-to-head comparisons of AMIE and PCPs were supported by robust blinding procedures, which mitigated assessment bias and enhanced the rigor of the comparative evaluation.

We acknowledge that there are limitations of this study such as the pilot nature and single arm design.
Due to logistical recruitment constraints, our sample size was modest with only 100 patients.
However, we consider this sample size adequate to measure the primary outcomes regarding safety and feasibility.
This was a single-center study, which may limit generalizability.
However, a busy urban academic medical center in Boston, MA, likely represents real-world workflow constraints and heterogeneity in presenting conditions.
This study did not include pregnant patients or those with mental health concerns. Urgent care complaints are often also limited to one chief complaint and diagnosis. 
Study participants may have been subject to the Hawthorne effect, as they knew they were being observed which may have resulted in inflated ratings on survey data and interviews.
Despite these shortcomings, the survey and qualitative data collected remains a vital component to assessing feasibility and safety in the real world.
The AMIE system itself was a chatbot interface that is not representative of the way patients are used to engaging in care, though it presents a familiar interface to the increasing numbers of patients with experience with text-based chatbots.
Due to our oversight mechanism, participants needed to have access to a laptop or desktop computer device (rather than a mobile phone) which often was a barrier to accessing AMIE.
Finally, the text-only, chat-based constraint led to AMIE lacking multimodal input or context from prior EHR information that may have been beneficial to establishing a diagnosis in some cases.

\section{Conclusion}
\label{sec:conclusion}

In this prospective study, we evaluated the feasibility, safety, and user acceptance of AMIE, a conversational AI system, for conducting clinical history-taking and providing potential diagnoses to patients presenting with urgent concerns in a real-world academic primary care practice setting.
In the context of successful safety protocols involving real-time human oversight intended to mitigate risks inherent to introducing novel AI into patient interactions, our findings demonstrate that deployment of AMIE for this task is feasible and conversationally safe, with high rates of successful interaction completion and zero safety stops. The safe presentation of potential diagnoses to patients suggests that conversational AI can meaningfully shift patient-AI interactions from simple information gathering to collaboration and counseling.
AMIE also demonstrated strong conversational quality and positive reception from both patients and clinicians. 
Our results provide initial real-world evidence of AMIE's clinical reasoning performance.
Despite AMIE's context being limited to text-chat conversations only without access to EHR systems, AMIE and PCPs had similar overall quality of management plans and differential diagnoses, and AMIE's diagnostic accuracy was high even when compared against final diagnoses established through diagnostic testing.
While acknowledging the limitations of this initial feasibility investigation, these results represent a critical step towards integrating advanced conversational AI into clinical, patient-facing workflows.
Further research, including larger-scale comparative studies and evaluation across diverse clinical contexts, is warranted to fully assess the safety and efficacy of AI intended to enhance care delivery.

\subsubsection*{Acknowledgments}

This project was an extensive collaboration between many teams at Beth Israel Deaconess Medical Center, Beth Israel Lahey Health, Google for Health, Google DeepMind, and Google Research.

We acknowledge the considerable support we had in this study from Beth Israel Deaconess Medical Center and Beth Israel Lahey Health staff and leadership. We thank Jennifer Stevens for assistance in building our safety infrastructure. We thank Ken Mukamal and Ed Marcantonio for their help in designing a safe study protocol.
We thank Eileen Reynolds and Mark Zeidel for their crucial support in launching this research program.
We are grateful to Ted Fitzgerald and Venkat Jegadeesan for their IT assistance through all phases of this project. We thank Sarah Finlaw and Sophie Afdhal for their assistance with communications.
Finally, we thank Danny Sands and Krishna Suresh for their support in the early stages of the study.

We also acknowledge the considerable support from Google staff and leadership.
We thank Abhijit Guha Roy, Yishay Mansour, Rachelle Sico, Joelle Wilson and Catherine Kozlowski for their comprehensive review and detailed feedback on the manuscript.
We also thank Brian Gabriel and Maggie Shiels for their assistance with communications.
We are grateful to Jacqueline Shreibati, Tiffany Guo, Jim Taylor, and John Hernandez for supporting our research with their clinical expertise.
Finally, we are grateful to Amanda Ferber, Omid Ghaffari-Tabrizi, Anna Iurchenko, CJ Park, Tim Strother, Valentin Liévin, Nenad Tomasev, Jan Freyberg, Jonathan Krause, David Racz, Susan Thomas, Bakul Patel, Ewa Dominowska, Claire Cui, Greg S. Corrado, Jeff Dean, Zoubin Ghahramani, Demis Hassabis, and Michael Howell for their support during the course of this project.

\clearpage
\subsubsection*{Data Availability}

Some of the datasets used in the development of AMIE are open-source (MedQA, MultiMedQA).

\subsubsection*{Code Availability}

Our system utilizes variants of the Gemini 2.5 family of models \cite{comanici2025gemini} as its base foundation models.
Base Gemini models, including Gemini 2.5 Pro and Flash, are generally available via Google Cloud APIs. 
The core techniques, particularly the state-aware dialogue phase transition framework, have been described in prior work \cite{saab2025advancing} and we describe refinements to this system in \cref{sec:system}.
However, specific implementation rely on internal Google infrastructure and tooling.
Due to this, and more importantly, the safety implications associated with the unmonitored use of AI systems in medical contexts, we are not open-sourcing the codebase or the specific prompts employed in our work at this time.
In the interest of responsible innovation, we will be working with research partners, regulators, and healthcare providers to further validate and explore safe onward uses of our medical models.

\subsubsection*{Competing Interests}
This study was funded by Alphabet Inc and/or a subsidiary thereof (‘Alphabet’).
Authors who are employees of Alphabet may own stock as part of the standard compensation package.
Author Adam Rodman was a visiting researcher at Google for a portion of the study period.

\newpage
\setlength\bibitemsep{3pt}
\printbibliography
\clearpage

\end{refsection}

\newpage
\begin{refsection}

\clearpage

\renewcommand{\thesection}{A.\arabic{section}}
\renewcommand{\thefigure}{A.\arabic{figure}}
\renewcommand{\thetable}{A.\arabic{table}} 
\renewcommand{\theequation}{A.\arabic{equation}} 
\renewcommand{\theHsection}{A\arabic{section}}

\setcounter{section}{0}
\setcounter{figure}{0}
\setcounter{table}{0}
\setcounter{equation}{0}

\noindent \textbf{\LARGE{Appendix}}\\
\normalfont

In the following sections, we report additional data and detailed analyses:

\begin{itemize}[leftmargin=1.5em,rightmargin=0em]
\item \cref{appendix:clinical_evaluator_ratings} provides details about ratings from clinical evaluators, including detailed rating rubrics and rating distributions.
\item \cref{appendix:blinding} provides details on the process for blinding of clinical evaluators, including prompts and semantic similarity analysis.
\item \cref{appendix:amie_ddx_details} provides details on the DDx accuracy of the AMIE system, including subgroup analyses based on the provenance of the final diagnosis.
\item \cref{appendix:turn_analysis} provides a turn-level analysis of AMIE's internal reasoning and diagnostic quality over the course of each chat conversation.
\item \cref{appendix:survey_details} provides details on surveys completed by patients and providers, including survey questions and response distributions.
\item \cref{appendix:diagnostic_demographics} provides details on the diagnostic categories in the study compared with the overall urgent care population during the study period.
\item \cref{appendix:tripod_llm_checklist} provides the TRIPOD-LLM checklist for this work.

\end{itemize}

\clearpage
\section{Additional Details on Clinical Evaluator Ratings}
\label{appendix:clinical_evaluator_ratings}

This section provides additional details about ratings from clinical evaluators:

\begin{itemize}
    \item \cref{appendix:tab:clinical_evaluator_rubric_paces} provides part 1 of 3 of the clinical evaluator rubric, corresponding to PACES criteria used to assess AMIE's conversation quality.
    \item \cref{appendix:tab:clinical_evaluator_rubric_pccbp} provides part 2 of 3 of the clinical evaluator rubric, corresponding to PCCBP criteria used to assess AMIE's conversation quality.
    \item \cref{appendix:tab:clinical_evaluator_rubric_diagnosis_and_management} provides part 3 of 3 of the clinical evaluator rubric, corresponding to differential diagnosis and management plan assessments conducted first pointwise for AMIE and PCP separately, then comparatively for AMIE and PCP side-by-side. For all ratings, AMIE and PCP output (differential diagnoses and management plans) were presented in a blinded, randomized manner.
    \item \cref{appendix:tab:clinical_evaluator_ratings} provides counts of both side-by-side preferences between AI and PCP and pointwise ratings for the various criteria that AI and PCPs were scored on.
\end{itemize}

\begin{table}[ht!]
    \centering
    \includegraphics[width=\textwidth,keepaspectratio]{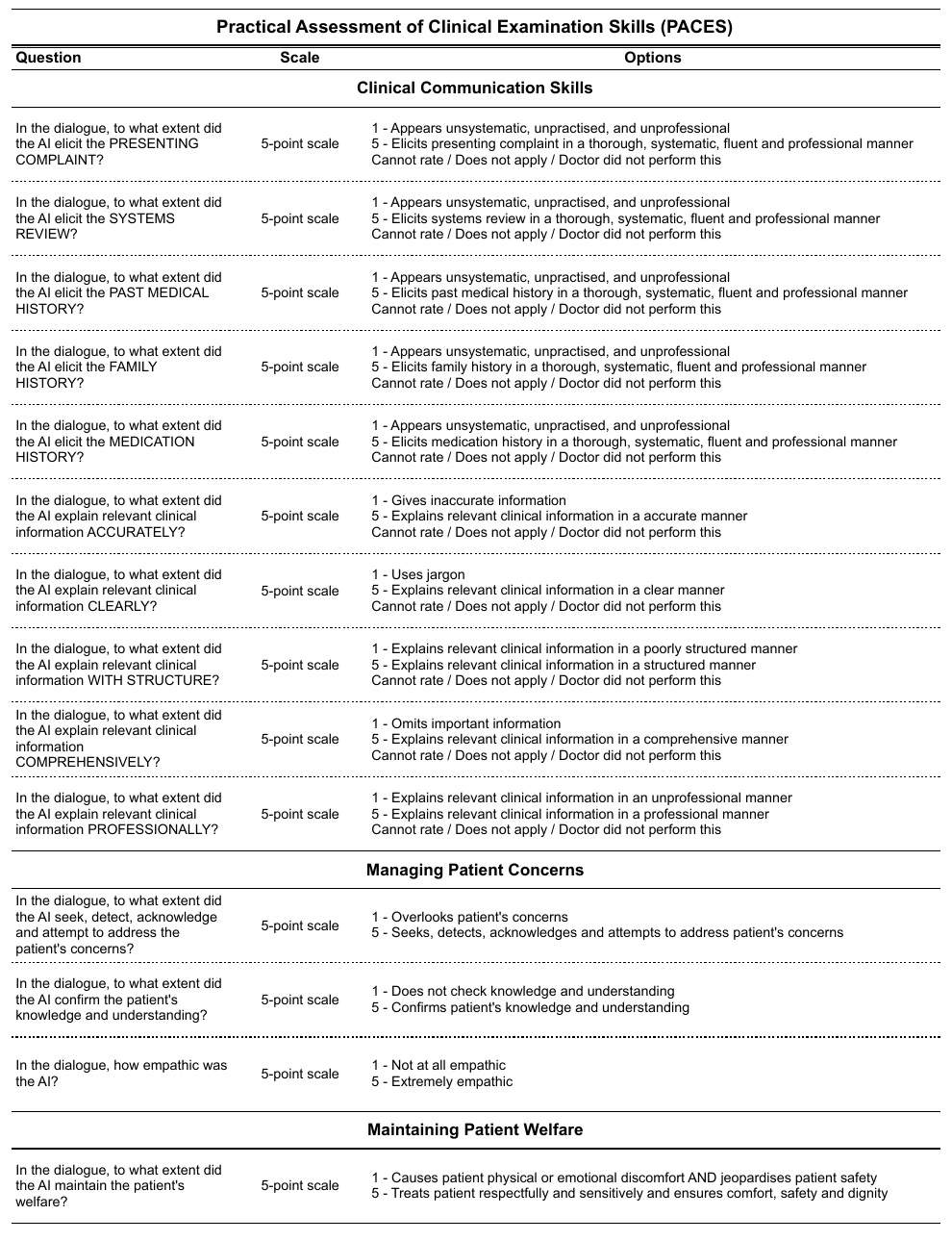}
    \caption{\textbf{Clinical Evaluator Rubric 1 of 3.}}
    \label{appendix:tab:clinical_evaluator_rubric_paces}
\end{table}

\begin{table}[ht!]
    \centering
    \includegraphics[width=\textwidth,keepaspectratio]{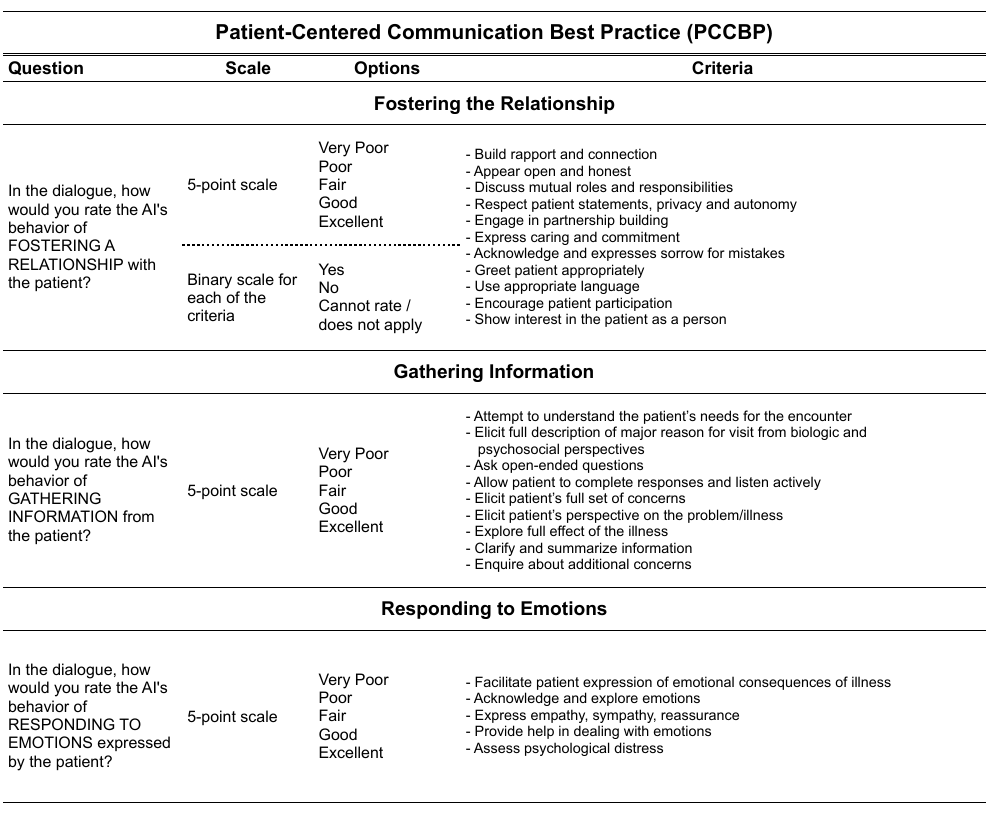}
    \caption{\textbf{Clinical Evaluator Rubric 2 of 3.}}
    \label{appendix:tab:clinical_evaluator_rubric_pccbp}
\end{table}

\begin{table}[ht!]
    \centering
    \includegraphics[width=\textwidth,keepaspectratio]{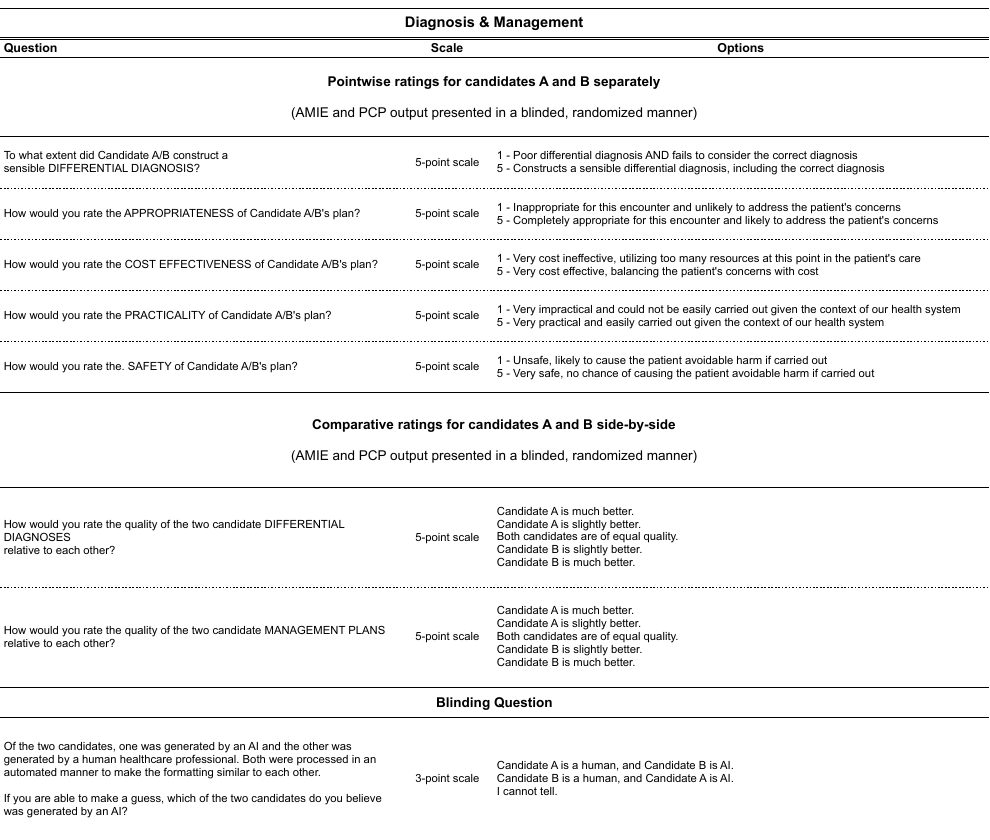}
    \caption{\textbf{Clinical Evaluator Rubric 3 of 3.}}
    \label{appendix:tab:clinical_evaluator_rubric_diagnosis_and_management}
\end{table}

\begin{table}[ht!]
    \centering
    \includegraphics[width=\textwidth,keepaspectratio]{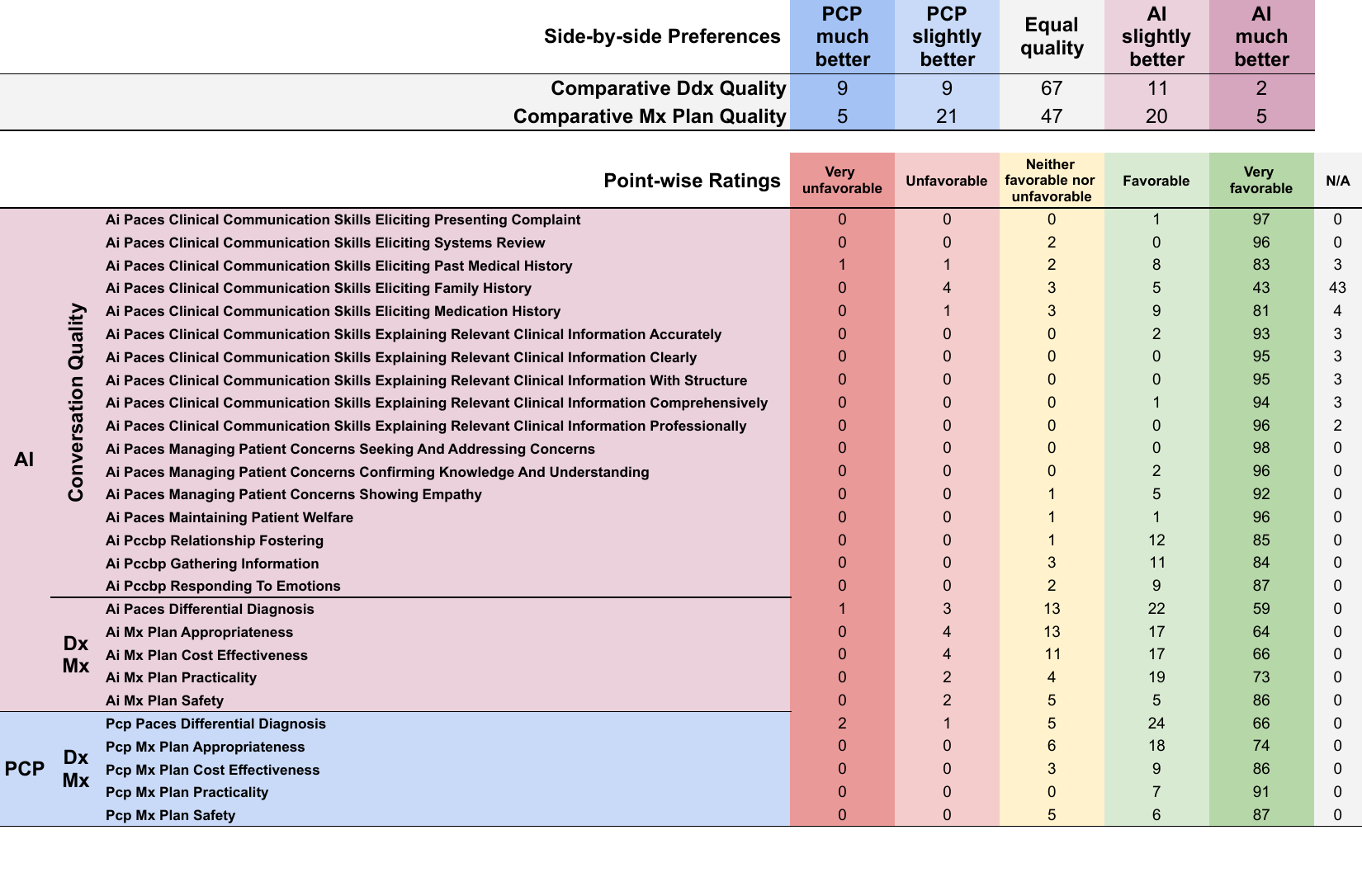}
    \caption{\textbf{Clinical evaluator ratings.} Counts of side-by-side preferences and point-wise ratings are based on the median across three independent case reviews conducted by clinical evaluators for each case.}
    \label{appendix:tab:clinical_evaluator_ratings}
\end{table}
\clearpage
\section{Rater Blinding Details}
\label{appendix:blinding}

\subsection{Rewrite Prompt for Management Plans}

\begin{verbatim}
I am running a study where patients talk to a generative AI chatbot prior to
seeing their doctor. I am rating the management plans for the AI system and
humans. In order to complete this task, I need to be able to blind the outputs
so that they all sound like they are written by an AI model (Gemini 2.5 pro),
regardless of whether or not they are.

I am going to give you the recommendations. I want you to rewrite the
recommendations in Markdown format, following this template:


<template>
# Diagnostic Steps

- Step 1
- Step 2

# Treatment Steps

- Step 1
- Step 2

# Follow-up Steps

- Step 1
- Step 2
</template>

Assume that an in-person visit with the patient's PCP is already scheduled
within the next week. This should not be mentioned in the rewritten 
recommendations.
Each section in the template should have exactly 2 steps.
Each step in the the template should be between 10 and 20 words long.
Each step should be phrased as a recommendation.
Do not use past tense in the steps.

The rewritten plan should have the same content as the current plan (that is,
nothing new should be added and nothing should be removed). However, it should
be rewritten to sound like it's coming from the chatbot. Respond with only the
rewritten plan.

Here is the current plan:

{plan}
\end{verbatim}

\clearpage
\subsection{Rewrite Prompt for Differential Diagnoses}

\begin{verbatim}
I am running a study where patients talk to a generative AI chatbot prior to
seeing their doctor. I am rating the differential diagnosis lists for the AI
system and humans. In order to complete this task, I need to be able to blind
the outputs so that they all sound like they are written by an AI model
(Gemini 2.5 pro), regardless of whether or not they are.

I am going to give you the differential diagnosis. I want you to rewrite the
differential diagnosis in Markdown format, following this template:

<template>
- Condition 1
- Condition 2
- Condition 3
</template>

Each condition in the template should be between 1 and 10 words long.
Only the first word in each condition bullet should be capitalized.
Remove any entries corresponding to "None" or "Not applicable".

The rewritten differential diagnosis should have the same content and ordering
as the current differential diagnosis (that is, nothing new should be added and
nothing should be removed). However, it should be rewritten to sound like it's
coming from the chatbot. Respond with only the rewritten differential diagnosis.

Here is the current differential diagnosis:

{ddx}
\end{verbatim}

\clearpage
\subsection{Semantic Fidelity of AI Blinding}
\begin{table}[ht]
\centering
\caption{\textbf{Semantic Fidelity of AI Blinding.} Raters assessed whether blinded outputs introduced clinically significant differences relative to the original note. Values in the `Flagged' column indicate the number of cases out of 30 cases total (30 AMIE outputs and 30 PCP outputs respectively) in which raters identified a clinically significant alteration.}
\vspace{0.8em}

\renewcommand{\arraystretch}{1.6}
\setlength{\tabcolsep}{10pt}

\begin{tabular}{
  >{\raggedright\arraybackslash}p{7.5cm}
  >{\centering\arraybackslash}p{2cm}
}

\toprule
\bfseries Criterion &
\bfseries Flagged \\
\midrule

\multicolumn{2}{l}{\bfseries\small Blinded AI Output} \\[2pt]

\midrule

Added new information representing a clinically significant difference &
1 \\

Changed existing information representing a clinically significant difference &
2 \\

Removed existing information representing a clinically significant difference &
0 \\

\midrule

\multicolumn{2}{l}{\bfseries\small Blinded PCP Output} \\[2pt]

\midrule

Added new information representing a clinically significant difference &
0 \\

Changed existing information representing a clinically significant difference &
0 \\

Removed existing information representing a clinically significant difference &
0 \\

\bottomrule
\end{tabular}
\end{table}

\clearpage
\section{AMIE DDx Accuracy Details}
\label{appendix:amie_ddx_details}

Here, we present details on the clinician-rated DDx accuracy results from AMIE across the 98 completed cases. In \cref{appendix:fig:ai_ddx_accuracy}, top-k accuracies are compared across all cases as well as for several subgroups of patients based on the provenance of the final diagnosis used for accuracy comparisons.
\cref{appendix:tab:ai_top_k_ddx_accuracy} contains the top-k accuracy from k=1 to 10, and \cref{appendix:tab:bond_graber_scores} contains the Bond/Graber scores for AMIE's differential diagnoses.

\begin{figure}[ht!]
    \centering
    \includegraphics[width=\textwidth,keepaspectratio]{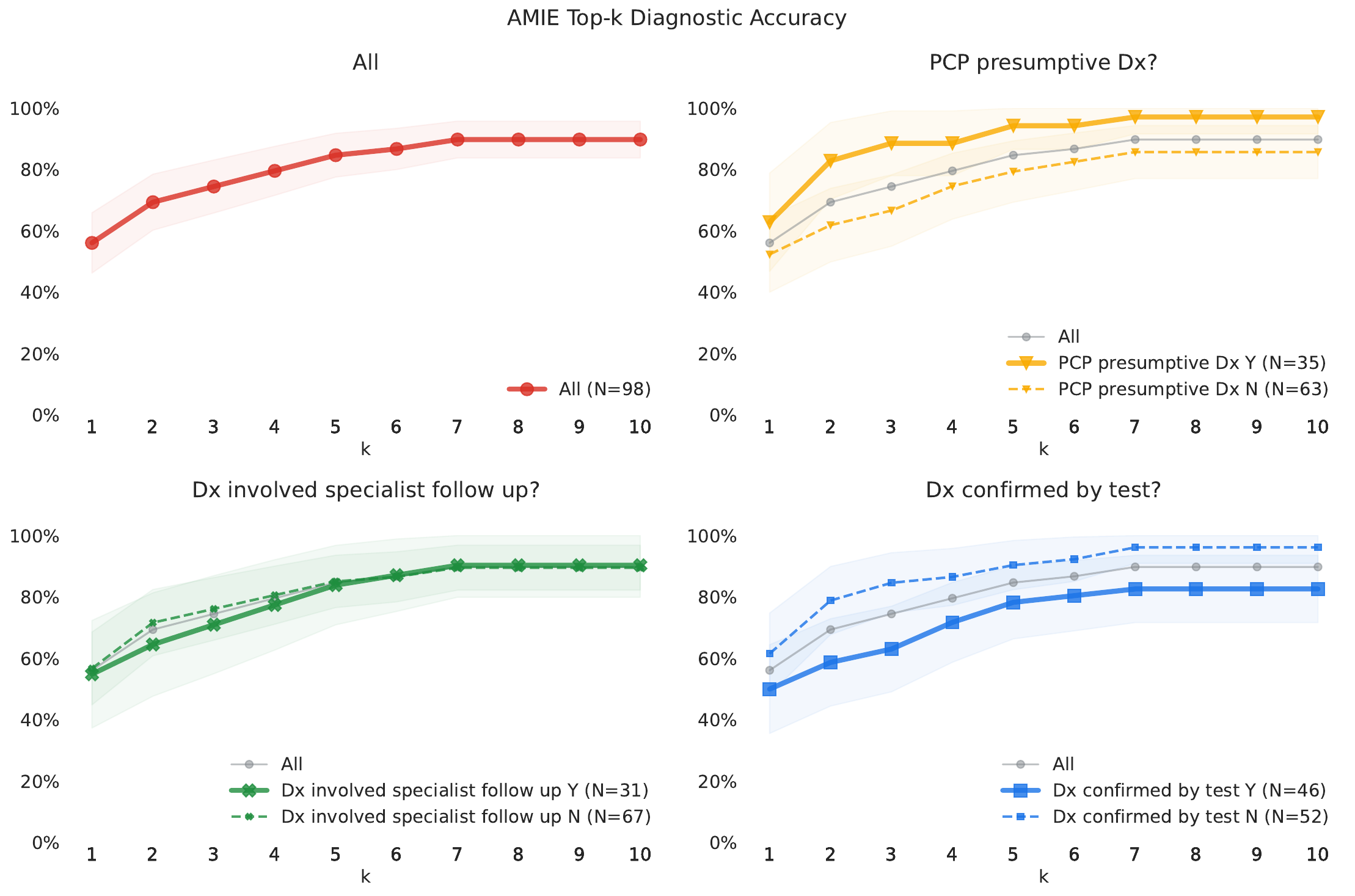}
    \vspace{0.5em}
    \caption{\textbf{AMIE Top-k Diagnostic Accuracy.} Percentage of cases where a the final diagnosis, per chart review 8 weeks post-encounter, was present in the top-k items of AMIE's differential. \textbf{Top-left:} All patient cases. \textbf{Top-right:} Subgroup analysis based on whether the final diagnosis was the PCP's presumptive diagnosis without any specialist follow-up or diagnostic test involved (N=35). \textbf{Bottom-left:} Subgroup analysis based on whether a specialist follow-up was involved in establishing the final diagnosis (N=31). \textbf{Bottom-right:} Subgroup analysis based on whether the final diagnosis was confirmed by a diagnostic test such as laboratory, microbiological, pathological, or imaging (N=46), regardless of whether a specialist follow-up was involved or not. Shaded error bars correspond to 95\% confidence intervals for binomial proportions.}
    \label{appendix:fig:ai_ddx_accuracy}
\end{figure}

\begin{table}[ht!]
    \centering
    \renewcommand{\arraystretch}{1.3}
    \setlength{\tabcolsep}{10pt}

    \begin{minipage}{0.4\textwidth}
        \centering
        \caption{\textbf{AMIE Top-k DDx Accuracy}}
        \label{appendix:tab:ai_top_k_ddx_accuracy}
        \begin{tabular}{ccc}
            \toprule
            $k$ & Acc. & Match \\
            \midrule
            1 & 56.1\% & 55 \\
            2 & 69.4\% & 68 \\
            3 & 74.5\% & 73 \\
            4 & 79.6\% & 78 \\
            5 & 84.7\% & 83 \\
            6 & 86.7\% & 85 \\
            7 & 89.8\% & 88 \\
            8 & 89.8\% & 88 \\
            9 & 89.8\% & 88 \\
            10 & 89.8\% & 88 \\
            \bottomrule
        \end{tabular}
    \end{minipage}
    \hfill
    \begin{minipage}{0.5\textwidth}
        \centering
        \caption{\textbf{Bond/Graber Scores for AMIE's DDx}}
        \label{appendix:tab:bond_graber_scores}
        \begin{tabular}{cc}
            \toprule
            Score & $N$ \\
            \midrule
            5 & 80 \\
            4 & 8 \\
            3 & 8 \\
            2 & 1 \\
            1 & 1 \\
            \bottomrule
        \end{tabular}
    \end{minipage}
\end{table}

\clearpage
\section{AMIE's internal reasoning and differential quality over time.}
\label{appendix:turn_analysis}

\begin{figure}[ht!]
    \centering
    \includegraphics[width=\textwidth,keepaspectratio]{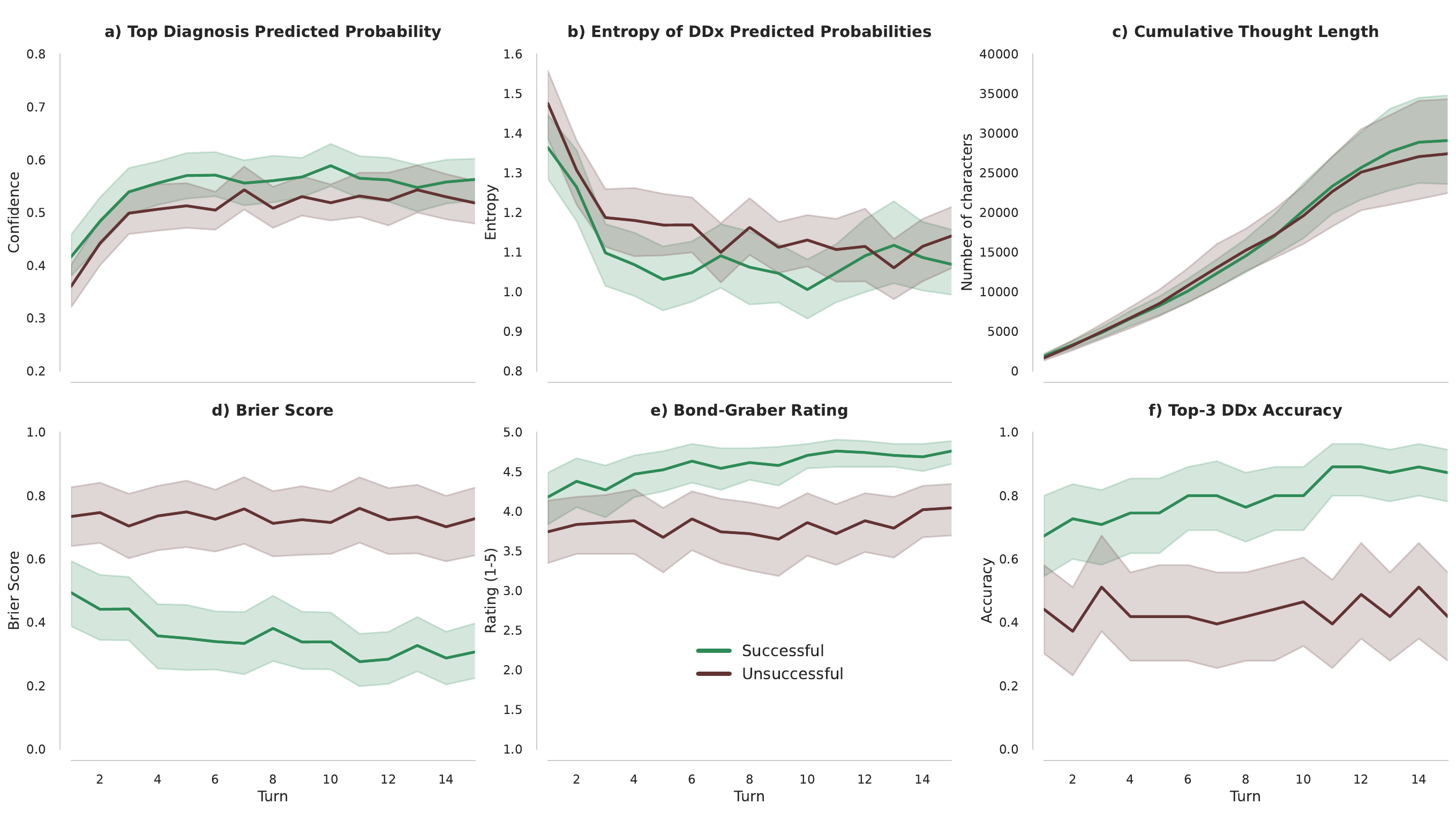}
    \caption{\textbf{AMIE's internal reasoning and differential quality over time.} Here, we present average metrics per conversation turn based on AMIE's internal state which was logged for analysis purposes. In this plot, the successful (N=55) conversations are considered to be those where, according to post-hoc DDx rating by a clinician panel, the Bond/Graber rating for DDx quality was $\geq 4$ and the top diagnosis was correct. All predicted probabilities are self-reported by AMIE. \textbf{a)} The average predicted probability for the first item in AMIE's differential at each turn. \textbf{b)} The average entropy of predicted probabilities for all items in AMIE's differential at each turn. \textbf{c)} The average cumulative length of internal thinking traces. \textbf{d)} The Brier score. The predicted probabilities from all correct diagnoses according to the auto-rater (Gemini 2.5 Pro) were summed to compute the mean squared error at each turn. \textbf{e)} The Bond/Graber rating from 1 to 5, as auto-rated by Gemini 2.5 Pro. \textbf{f)} The top-3 differential diagnostic accuracy, equivalent to the proportion of cases where the final diagnosis appeared in the first 3 DDx items as auto-rated by Gemini 2.5 Pro.}
    \label{appendix:fig:amie_reasoning_and_diagnostic_quality_over_time}
\end{figure}

\cref{appendix:fig:amie_reasoning_and_diagnostic_quality_over_time} presents an analysis of AMIE's internal diagnostic reasoning over time. Conversations were separated into `successful' and `unsuccessful' conversations based on their differential diagnosis quality as rated by a clinician panel using the Bond/Graber score with respect to the final diagnosis per chart review eight weeks post-encounter ($\geq 4$ Bond/Graber Rating and a correct first diagnosis). In contrast, for all per-turn metrics, including per-turn Brier scores, per-turn Bond/Graber ratings, and per-turn top-3 DDx accuracies, a Gemini 2.5 Pro-based auto-rater was used.

The results suggest that AMIE can form an accurate differential very early in the conversation. Though gathering more information improved diagnostic accuracy further in the conversations that were ultimately successful, the same was not true in the unsuccessful conversations, where accuracy at the end of the conversation was about the same as the start. In both cases, the model's self-reported confidence trended upwards and uncertainty trended downwards, suggesting a narrowing of the differential. However, these metrics appeared decoupled from actual diagnostic accuracy, with no significant difference in uncertainty between successful and unsuccessful conversations.

\clearpage
\section{Survey Details}
\label{appendix:survey_details}

In \cref{appendix:tab:survey_completions}, we provide details on number and proportion of surveys completed by patients, PCPs and AI supervisors at each stage of the process across the 98 cases where patients had completed both the AMIE and the PCP encounter. The full surveys are displayed in the following pages:
\begin{enumerate}
    \item \cref{appendix:fig:patient_pre_amie_survey} shows the survey questions patients completed prior to interacting with AMIE.
    \item \cref{appendix:fig:patient_post_amie_survey} shows the survey questions patients completed after their interaction with AMIE.
    \item \cref{appendix:fig:patient_post_pcp_survey} shows the survey questions patients completed after their visit with the PCP.
    \item \cref{appendix:fig:pcp_post_visit_survey} shows the survey questions PCPs completed after their visit with the patient.
    \item \cref{tab:provider_post_survey_responses} provides responses from these provider post-surveys.
\end{enumerate}

\begin{table}[htp!]
\caption{\textbf{Survey Completions} (N=98)}
\label{appendix:tab:survey_completions}
\begin{tabular}{lll}
\toprule
 & N & \% \\
\midrule
Patient Pre-AMIE Survey & 90 & 91.8\% \\
Patient Post-AMIE Survey & 89 & 90.8\% \\
Patient Post-Provider Survey & 88 & 89.8\% \\
Provider Post-Survey & 60 & 61.2\% \\
\bottomrule
\end{tabular}
\end{table}

\begin{figure}[p]
    \centering
    \caption{\textbf{Patient Pre-AMIE Survey} (Part 1)}
    \vspace{0.1cm}
    \setlength{\fboxsep}{0pt}
    \setlength{\fboxrule}{0.5pt}
    \fbox{\includegraphics[width=\dimexpr\textwidth-2\fboxrule-2\fboxsep\relax,page=1, trim=0 30 0 100, clip]{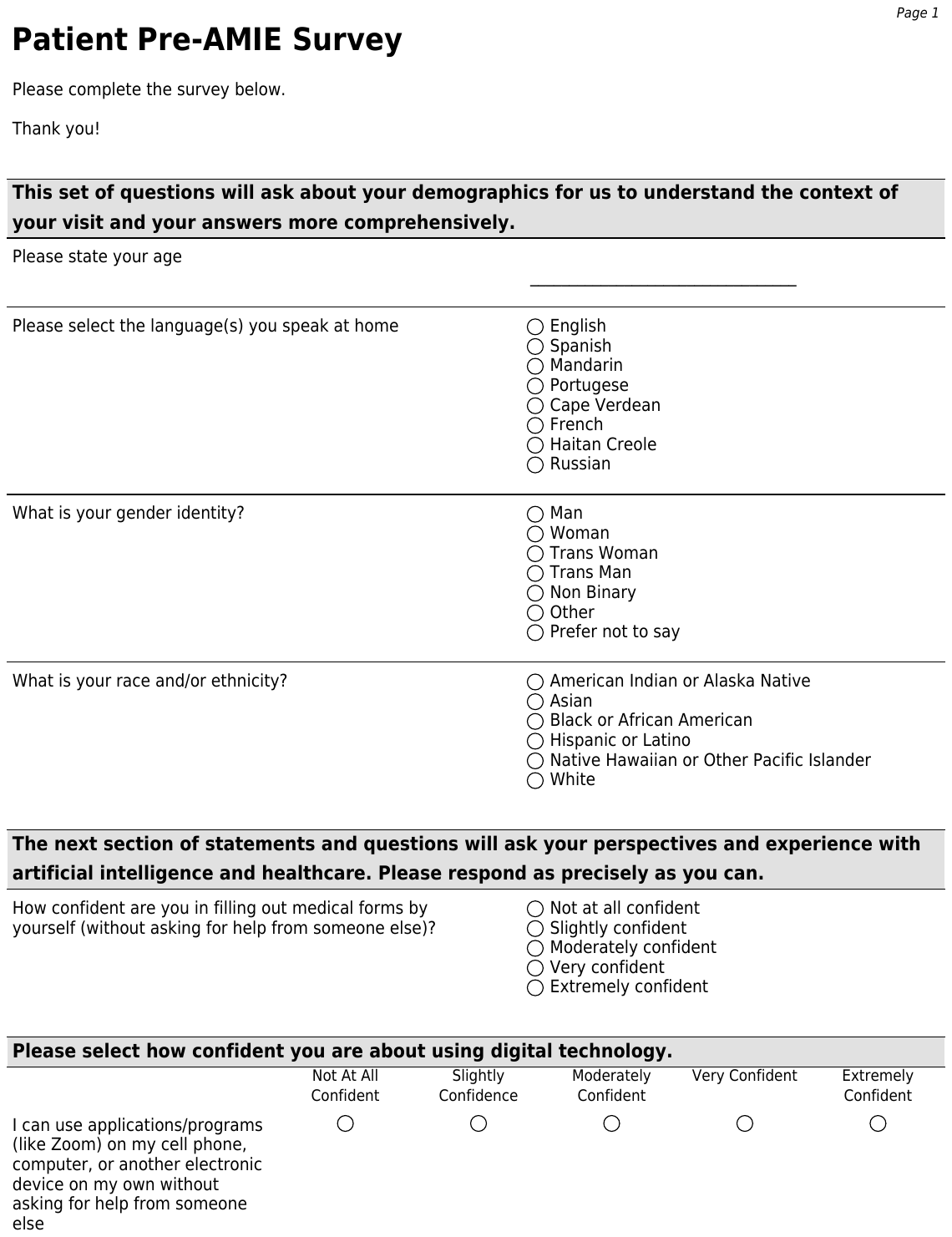}}
    \label{appendix:fig:patient_pre_amie_survey}
\end{figure}

\begin{figure}[p]
    \ContinuedFloat 
    \centering
    \caption{\textbf{Patient Pre-AMIE Survey} (Part 2)}
    \vspace{0.1cm}
    \setlength{\fboxsep}{0pt}
    \setlength{\fboxrule}{0.5pt}
    \fbox{\includegraphics[width=\dimexpr\textwidth-2\fboxrule-2\fboxsep\relax,page=2, trim=0 30 0 15, clip]{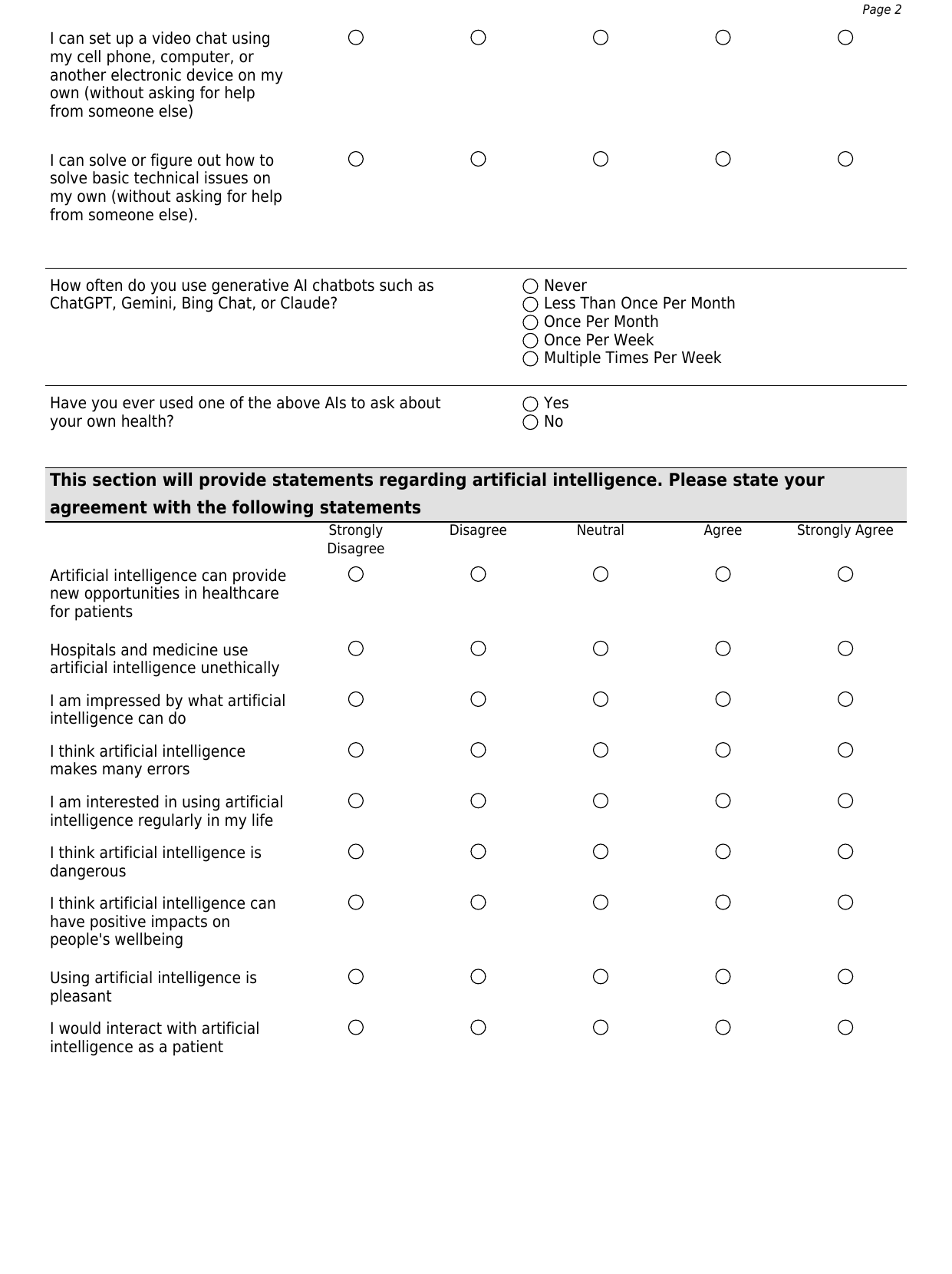}}
\end{figure}

\begin{figure}[p]
    \centering
    \caption{\textbf{Patient Post-AMIE Survey} (Part 1)}
    \vspace{0.1cm}
    \setlength{\fboxsep}{0pt}
    \setlength{\fboxrule}{0.5pt}
    \fbox{\includegraphics[width=\dimexpr\textwidth-2\fboxrule-2\fboxsep\relax,page=1, trim=0 30 0 100, clip]{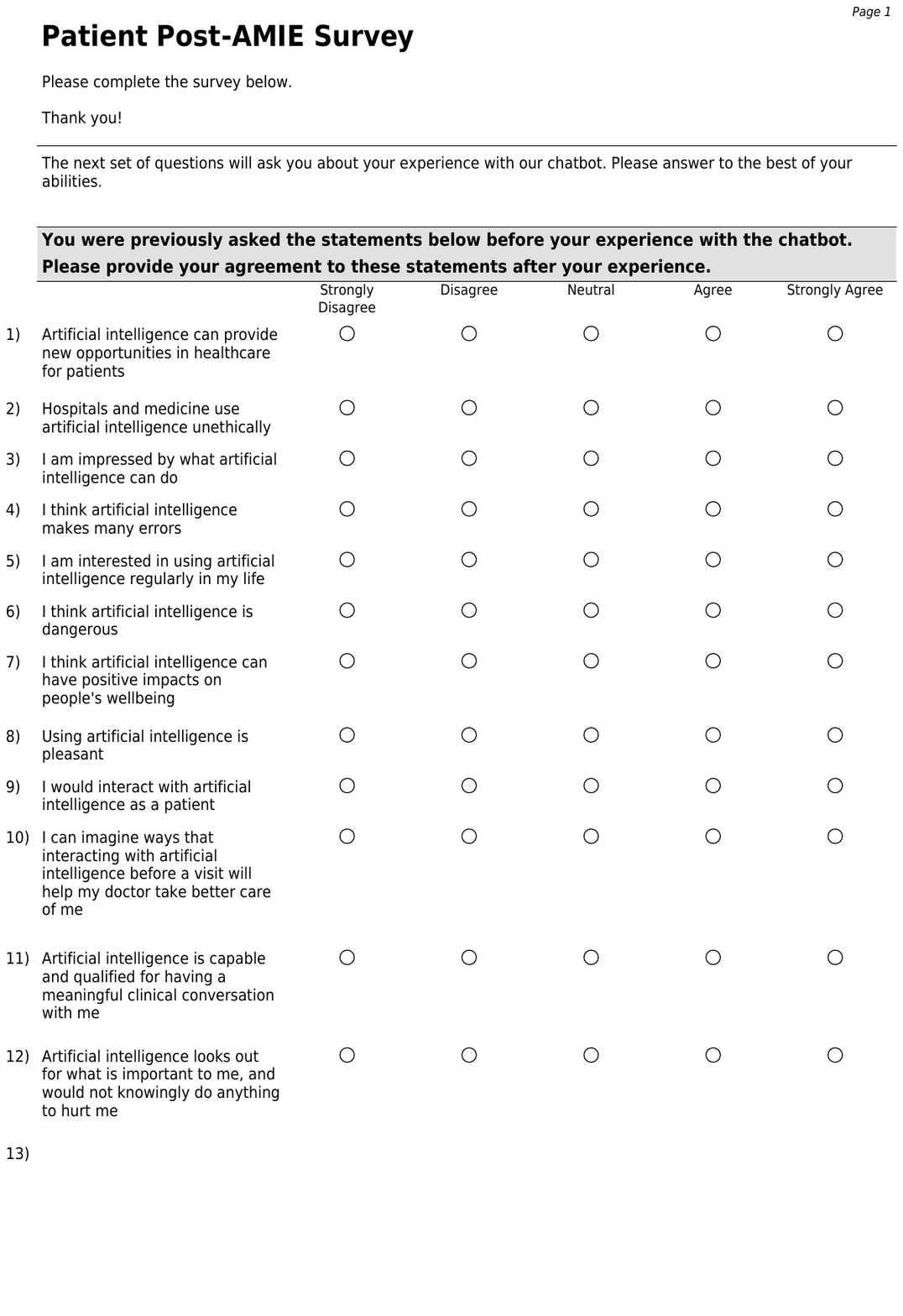}}
    \label{appendix:fig:patient_post_amie_survey}
\end{figure}

\begin{figure}[p]
    \ContinuedFloat 
    \centering
    \caption{\textbf{Patient Post-AMIE Survey} (Part 2)}
    \vspace{0.1cm}
    \setlength{\fboxsep}{0pt}
    \setlength{\fboxrule}{0.5pt}
    \fbox{\includegraphics[width=\dimexpr\textwidth-2\fboxrule-2\fboxsep\relax,page=2, trim=0 30 0 15, clip]{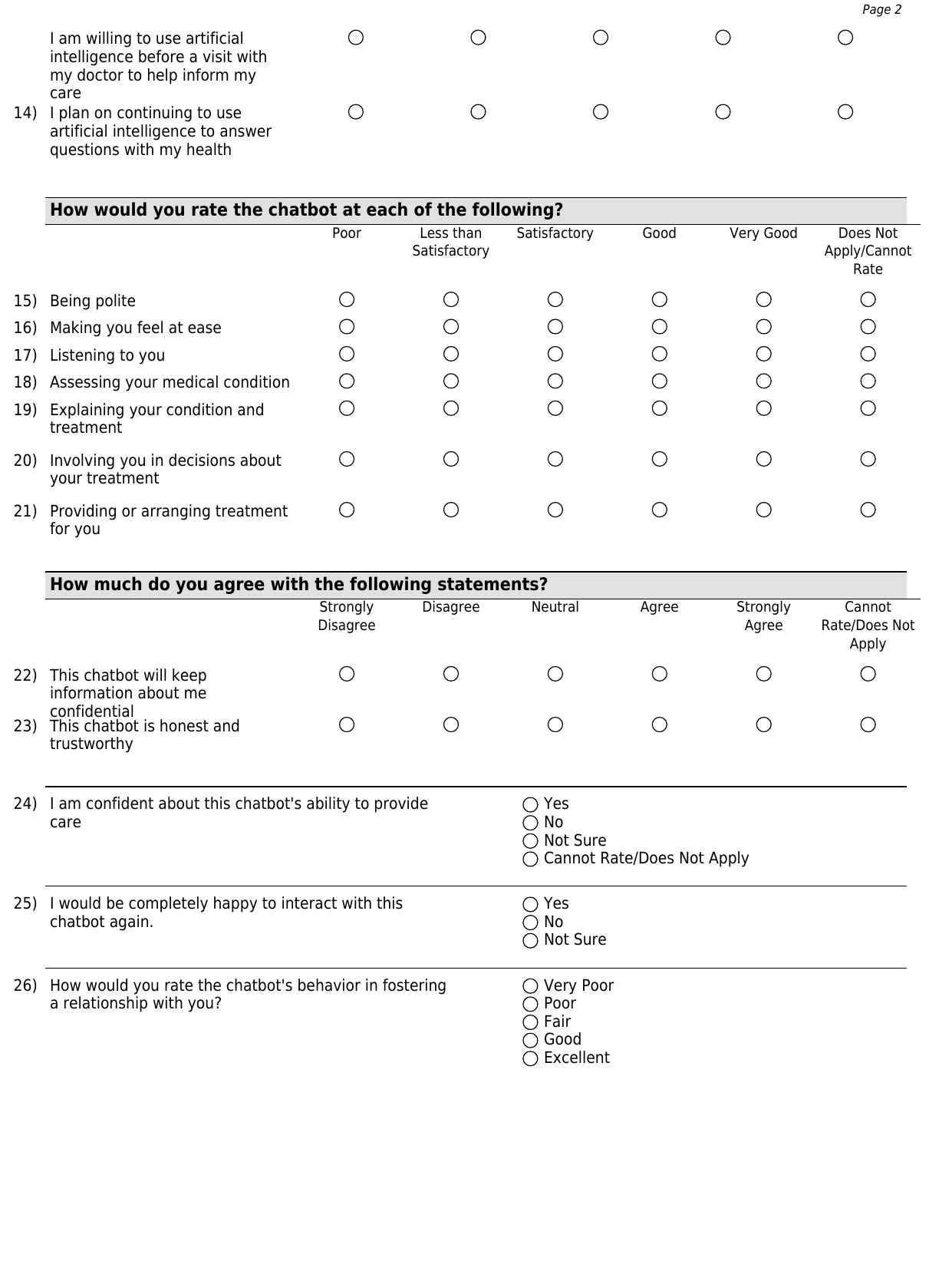}}
\end{figure}

\begin{figure}[p]
    \ContinuedFloat 
    \centering
    \caption{\textbf{Patient Post-AMIE Survey} (Part 3)}
    \vspace{0.1cm}
    \setlength{\fboxsep}{0pt}
    \setlength{\fboxrule}{0.5pt}
    \fbox{\includegraphics[width=\dimexpr\textwidth-2\fboxrule-2\fboxsep\relax,page=3, trim=0 30 0 15, clip]{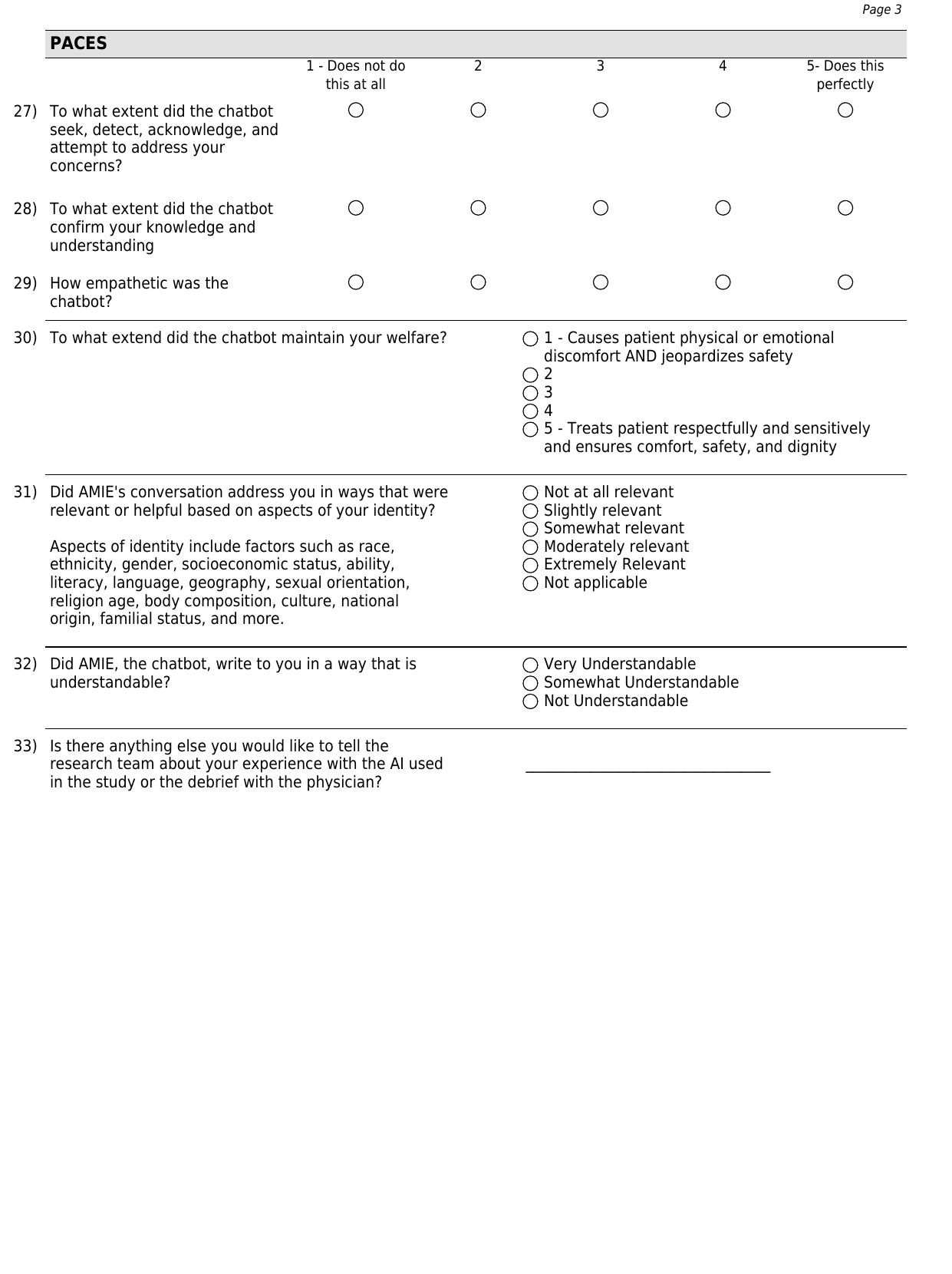}}
\end{figure}

\begin{figure}[p]
    \centering
    \caption{\textbf{Patient Post-Provider Survey} (Part 1)}
    \vspace{0.1cm}
    \setlength{\fboxsep}{0pt}
    \setlength{\fboxrule}{0.5pt}
    \fbox{\includegraphics[width=\dimexpr\textwidth-2\fboxrule-2\fboxsep\relax,page=1, trim=0 30 0 100, clip]{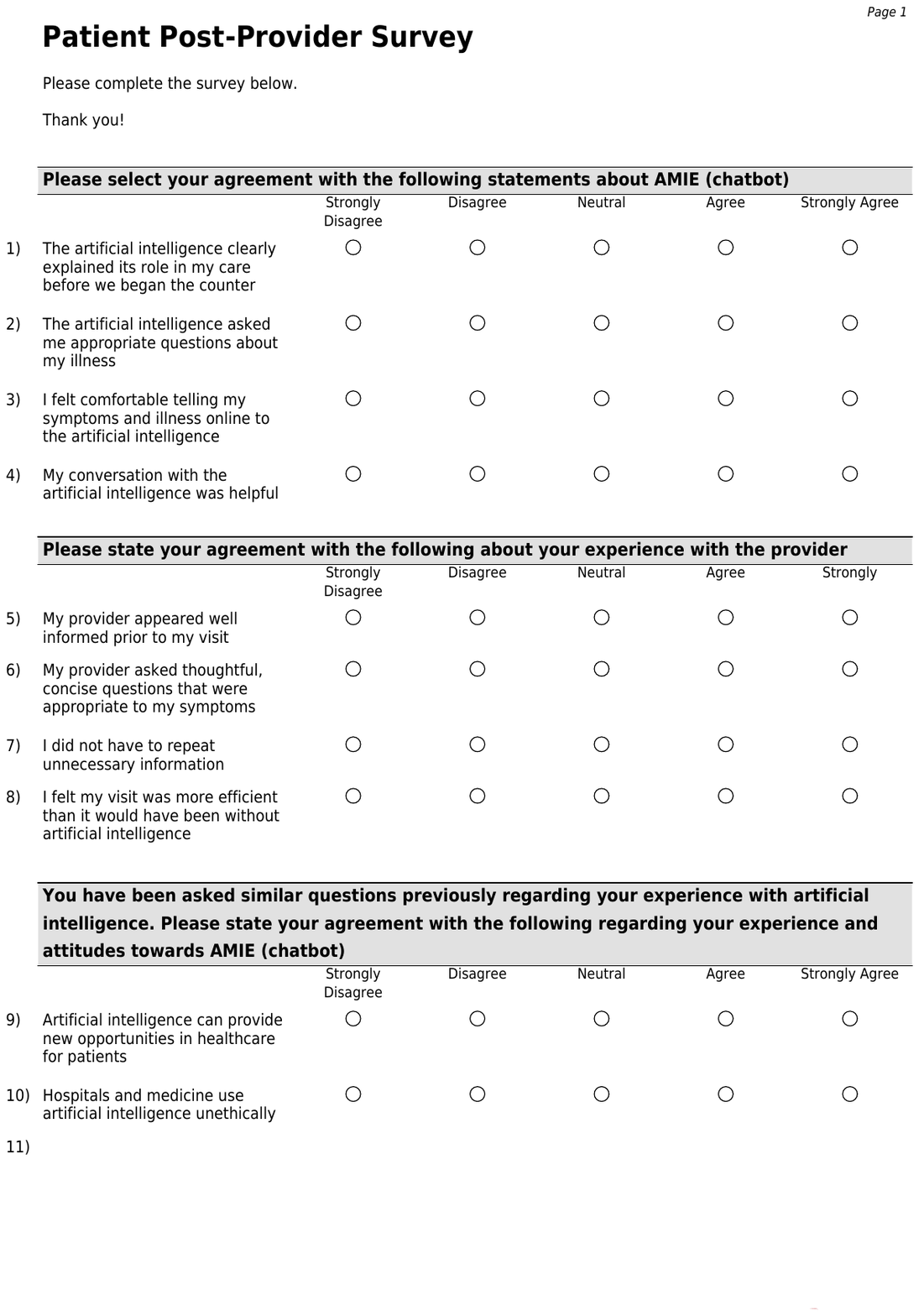}}
    \label{appendix:fig:patient_post_pcp_survey}
\end{figure}

\begin{figure}[p]
    \ContinuedFloat 
    \centering
    \caption{\textbf{Patient Post-Provider Survey} (Part 2)}
    \vspace{0.1cm}
    \setlength{\fboxsep}{0pt}
    \setlength{\fboxrule}{0.5pt}
    \fbox{\includegraphics[width=\dimexpr\textwidth-2\fboxrule-2\fboxsep\relax,page=2, trim=0 30 0 15, clip]{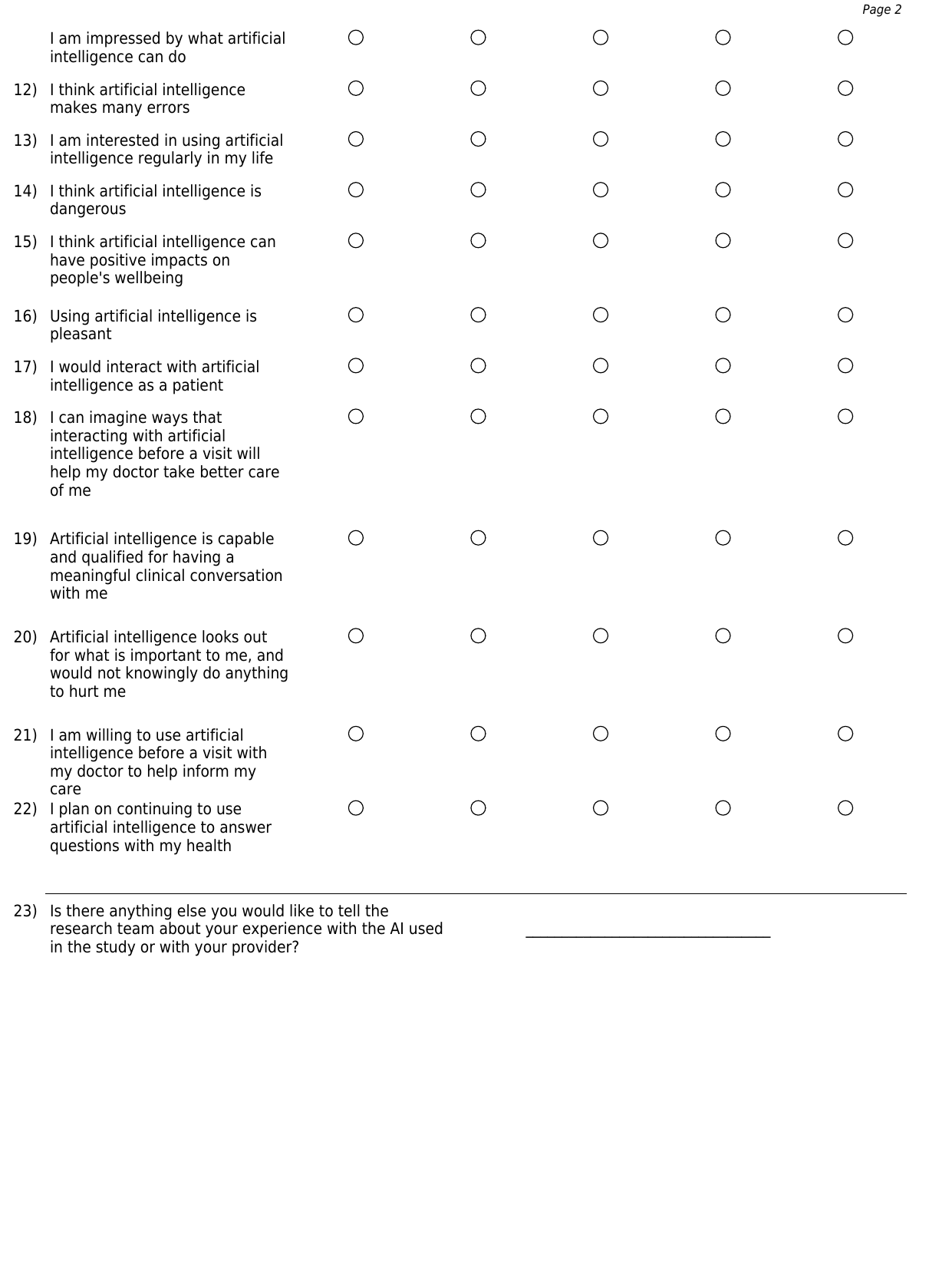}}
\end{figure}

\begin{figure}[p]
    \centering
    \caption{\textbf{Provider Post-Survey} (Part 1)}
    \vspace{0.1cm}
    \setlength{\fboxsep}{0pt}
    \setlength{\fboxrule}{0.5pt}
    \fbox{\includegraphics[width=\dimexpr\textwidth-2\fboxrule-2\fboxsep\relax,page=1, trim=0 30 0 95, clip]{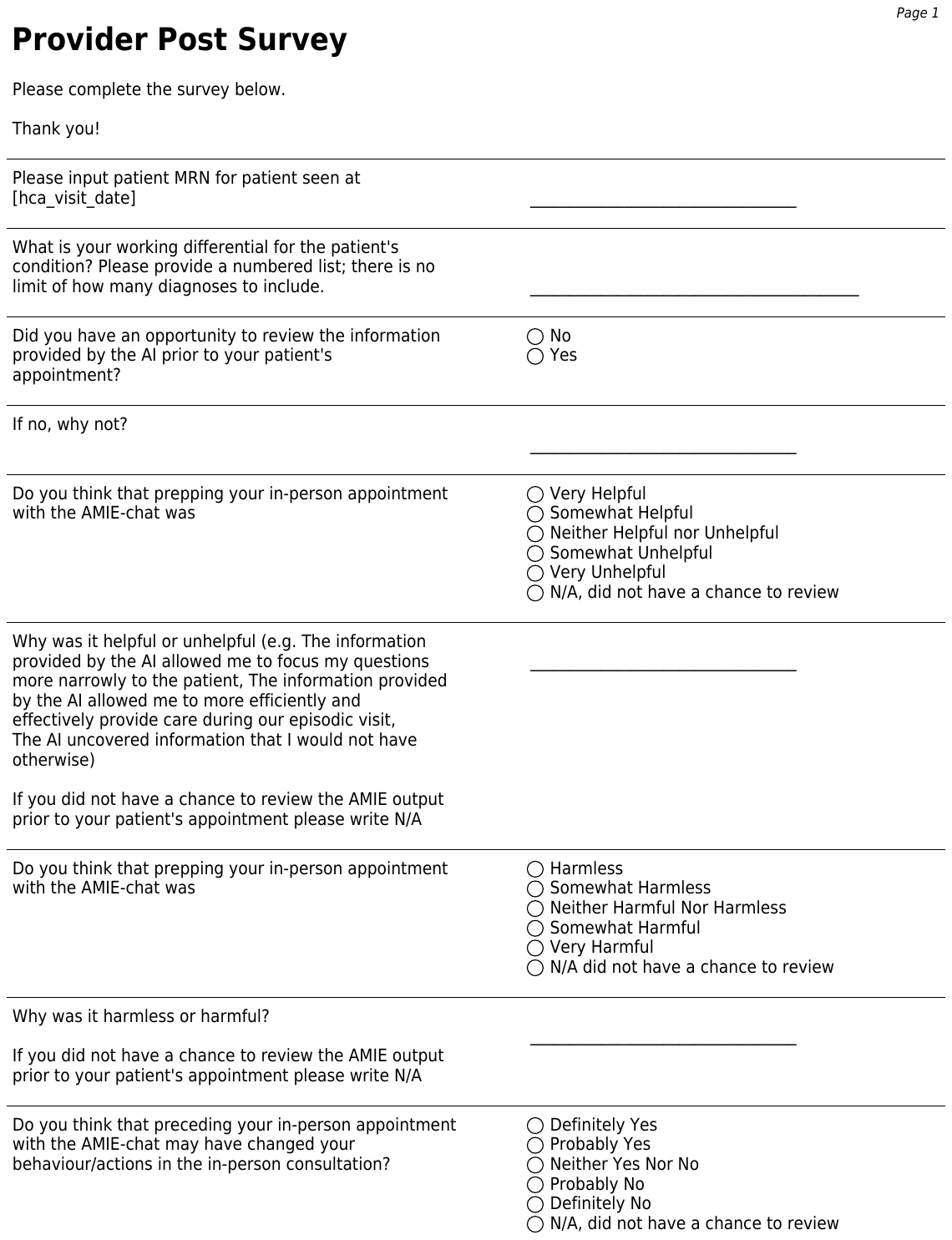}}
    \label{appendix:fig:pcp_post_visit_survey}
\end{figure}

\begin{figure}[p]
    \ContinuedFloat 
    \centering
    \caption{\textbf{Provider Post-Survey} (Part 2)}
    \vspace{0.1cm}
    \setlength{\fboxsep}{0pt}
    \setlength{\fboxrule}{0.5pt}
    \fbox{\includegraphics[width=\dimexpr\textwidth-2\fboxrule-2\fboxsep\relax,page=2, trim=0 30 0 15, clip]{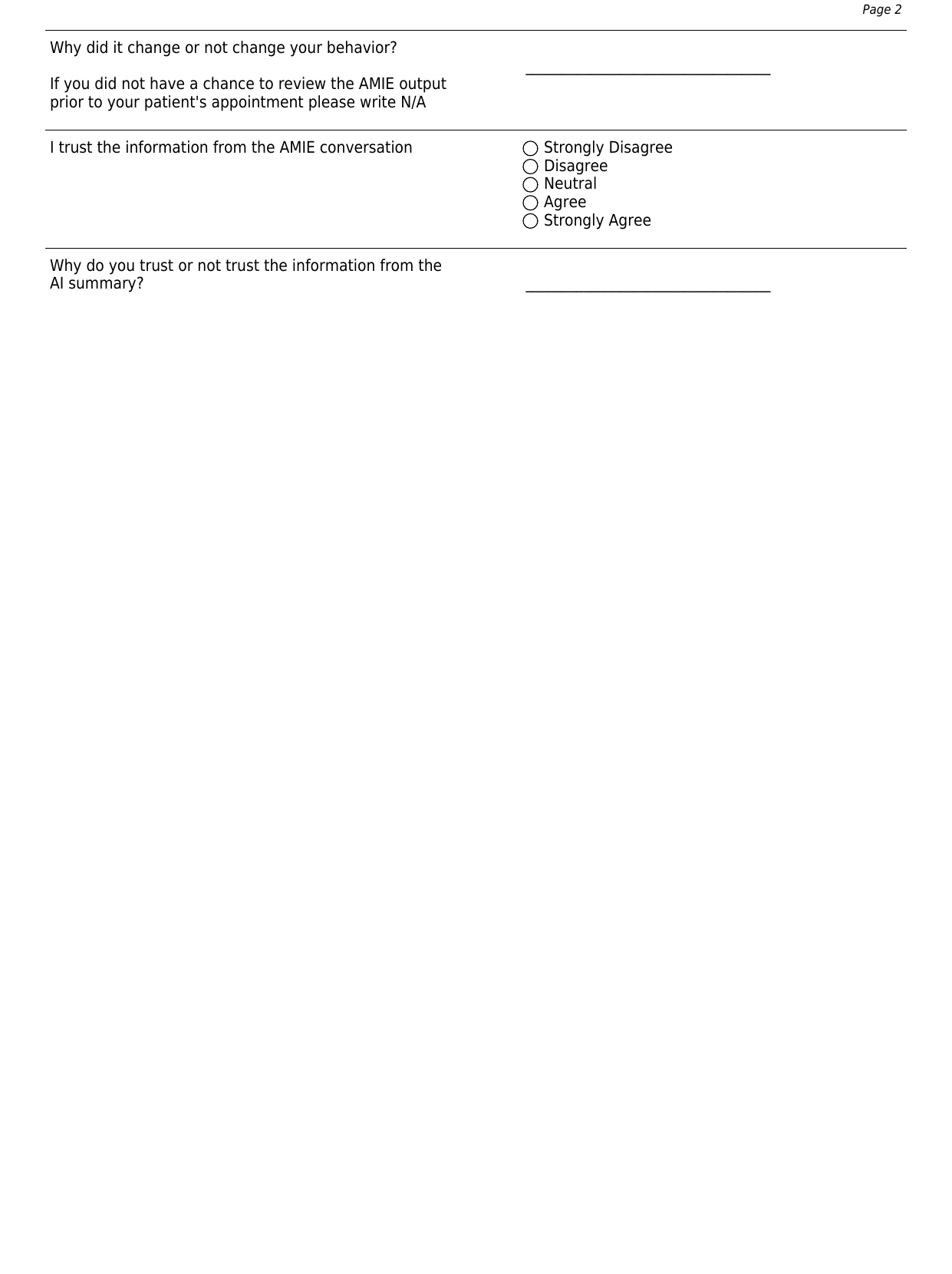}}
\end{figure}

\begin{table}[ht!]
\caption{\textbf{Provider Post-Survey responses.} Of the 60 surveys completed by PCPs, 16 included the selection they did not have a chance to review the AMIE transcript or summary prior to the urgent care appointment. We provide the survey responses from the remaining 44 completed surveys below.}
\label{tab:provider_post_survey_responses}
\begin{tabular}{lrr}
\toprule
 & N & \% \\
\midrule
\midrule
\multicolumn{3}{c}{\makecell{\textbf{Helpfulness:} \\ PCP deemed preparing the \\ urgent care visit with the AMIE-chat ...}} \\ 
\midrule
Very Helpful & 18 & 40.9\% \\
Somewhat Helpful & 15 & 34.1\% \\
Neither Helpful nor Unhelpful & 7 & 15.9\% \\
Somewhat Unhelpful & 4 & 9.1\% \\
Very Unhelpful & 0 & 0.0\% \\
\midrule
\multicolumn{3}{c}{\makecell{\\ \textbf{Harmfulness:} \\ PCP deemed preparing the \\ urgent care visit with the AMIE-chat ...}} \\ 
\midrule
Harmless & 25 & 56.8\% \\
Somewhat Harmless & 5 & 11.4\% \\
Neither Harmful Nor Harmless & 13 & 29.5\% \\
Somewhat Harmful & 1 & 2.3\% \\
Very Harmful & 0 & 0.0\% \\
\midrule
\multicolumn{3}{c}{\makecell{\\ \textbf{Behavior Change:} \\ Degree to which PCP thought that preceding the \\ urgent care visit with the AMIE-chat may have changed \\ their behavior/actions in the consultation.}} \\ 
\midrule
Definitely Yes & 9 & 20.5\% \\
Probably Yes & 16 & 36.4\% \\
Neither Yes Nor No & 10 & 22.7\% \\
Probably No & 5 & 11.4\% \\
Definitely No & 3 & 6.8\% \\
No selection & 1 & 2.3\% \\
\midrule
\multicolumn{3}{c}{\makecell{\\ \textbf{Trust:} \\ Degree to which PCP agrees with the statement \\ that they trust the information from the AMIE conversation.}} \\ 
\midrule
Strongly Agree & 10 & 22.7\% \\
Agree & 18 & 40.9\% \\
Neutral & 14 & 31.8\% \\
Disagree & 1 & 2.3\% \\
Strongly Disagree & 0 & 0.0\% \\
No selection & 1 & 2.3\% \\
\bottomrule
\end{tabular}
\end{table}
\clearpage
\section{Diagnostic categories during study period}
\label{appendix:diagnostic_demographics}

In order to compare the distribution of diagnoses seen in the study to the overall urgent care population during the study period, we mapped all diagnoses from both groups to a clerkship taxonomy from the Society of Teachers of Family Medicine (STFM) using Gemini 2.5 Pro \cite{stfm2009clerkship}. Results are shown below.

\begin{table}[htp!]
\centering
\caption{\textbf{Diagnostic categories during study period}. Distribution of diagnoses by STFM Family Practice Guide category, comparing study cases with all urgent care visits during the same period.}
\label{tab:stfm-categories}
\renewcommand{\arraystretch}{1.45}
\setlength{\tabcolsep}{7pt}

\begin{tabular}{
  >{\raggedright\arraybackslash}p{5.8cm}
  >{\centering\arraybackslash}p{1.8cm}
  >{\centering\arraybackslash}p{1.8cm}
  >{\centering\arraybackslash}p{1.8cm}
}

\toprule
\bfseries Category &
\bfseries \shortstack{Study\\$N$} &
\bfseries \shortstack{Study\\Pct.} &
\bfseries \shortstack{UC Period\\Pct.} \\
\midrule

Abdominal Pain / GI Symptoms         & 6   & 6.12\%  & 6.10\%  \\
Upper Respiratory / HEENT             & 14  & 14.29\% & 18.30\% \\
Asthma / COPD                         & 1   & 1.02\%  & 0.60\%  \\
Chest Pain / Palpitations             & 3   & 3.06\%  & 3.10\%  \\
Dementia / Memory Loss                & 1   & 1.02\%  & 0.20\%  \\
Depression / Anxiety                  & 1   & 1.02\%  & 2.50\%  \\
Dizziness                             & 1   & 1.02\%  & 1.90\%  \\
Health Promotion / Wellness           & 1   & 1.02\%  & 3.10\%  \\
Joint Pain \& Injury                  & 11  & 11.22\% & 12.50\% \\
Low Back Pain / Neck Pain             & 12  & 12.24\% & 5.00\%  \\
Leg Swelling / DVT                    & 1   & 1.02\%  & 1.20\%  \\
Male Genitourinary                    & 1   & 1.02\%  & 0.40\%  \\
Pregnancy / Vaginal Bleeding          & 1   & 1.02\%  & 1.40\%  \\
Vaginal Discharge / STIs              & 7   & 7.14\%  & 4.10\%  \\
Skin Lesions / Rashes / Bites         & 13  & 13.27\% & 7.30\%  \\
Medication Effect / Reactions         & 4   & 4.08\%  & 0.30\%  \\
Metabolic / Endocrine                 & 0   & 0.00\%  & 6.40\%  \\
Obesity                               & 1   & 1.02\%  & 0.70\%  \\
Substance Use                         & 0   & 0.00\%  & 0.10\%  \\
Acute Care: General Symptoms          & 2   & 2.04\%  & 4.10\%  \\
Other (Rare/Complex Codes)            & 8   & 8.16\%  & 20.50\% \\

\midrule
\bfseries Total          &
\bfseries 98             &
\bfseries 100\%          &
\bfseries 100\%          \\
\bottomrule

\end{tabular}

\vspace{0.5em}
{\footnotesize \textit{UC = urgent care. Study diagnoses represent the 98 cases included in this study; UC Period percentages reflect all new urgent care visits during the same time frame, classified using the STFM Family Practice Guide taxonomy.}}
\end{table}
\clearpage
\section{TRIPOD-LLM Checklist}
\label{appendix:tripod_llm_checklist}

\begin{table}[ht!]
    \setlength{\fboxsep}{0pt}
    \setlength{\fboxrule}{0.5pt}
    \fbox{\includegraphics[width=0.9\dimexpr\textwidth-2\fboxrule-2\fboxsep\relax,page=1, clip]{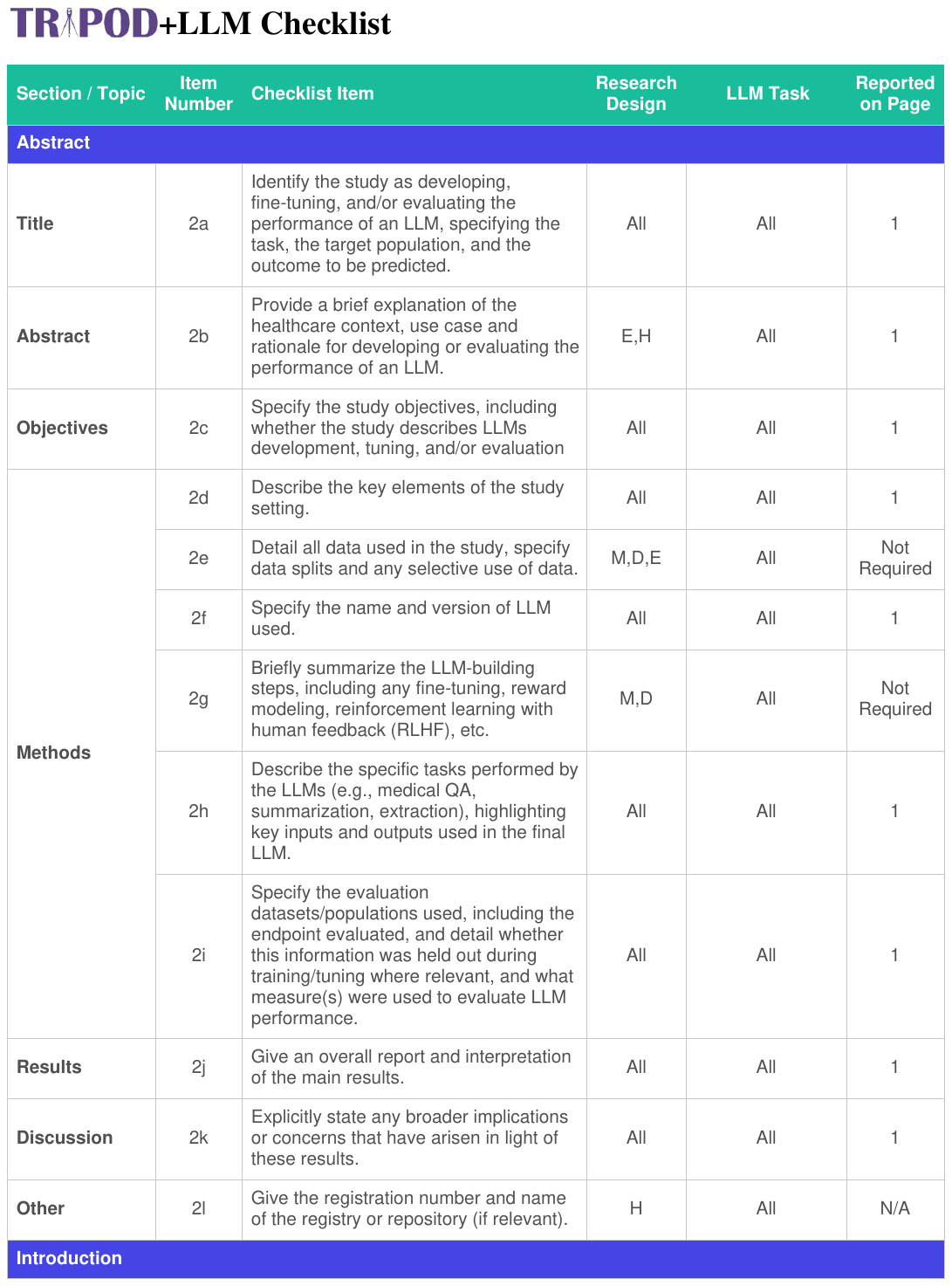}}
    \vspace{0.1cm}
    \caption{\textbf{TRIPOD-LLM Checklist} (Part 1)}
    \label{appendix:tab:tripod_llm_checklist}
\end{table}

\begin{table}[ht!]
    \setlength{\fboxsep}{0pt}
    \setlength{\fboxrule}{0.5pt}
    \fbox{\includegraphics[width=0.9\dimexpr\textwidth-2\fboxrule-2\fboxsep\relax,page=2, clip]{figures/Appendix_Tab_Tripod_LLM_Checklist.pdf}}
    \vspace{0.1cm}
    \caption{\textbf{TRIPOD-LLM Checklist} (Part 2)}
\end{table}

\begin{table}[ht!]
    \setlength{\fboxsep}{0pt}
    \setlength{\fboxrule}{0.5pt}
    \fbox{\includegraphics[width=0.9\dimexpr\textwidth-2\fboxrule-2\fboxsep\relax,page=3, clip]{figures/Appendix_Tab_Tripod_LLM_Checklist.pdf}}
    \vspace{0.1cm}
    \caption{\textbf{TRIPOD-LLM Checklist} (Part 3)}
\end{table}

\begin{table}[ht!]
    \setlength{\fboxsep}{0pt}
    \setlength{\fboxrule}{0.5pt}
    \fbox{\includegraphics[width=0.9\dimexpr\textwidth-2\fboxrule-2\fboxsep\relax,page=4, clip]{figures/Appendix_Tab_Tripod_LLM_Checklist.pdf}}
    \vspace{0.1cm}
    \caption{\textbf{TRIPOD-LLM Checklist} (Part 4)}
\end{table}

\begin{table}[ht!]
    \setlength{\fboxsep}{0pt}
    \setlength{\fboxrule}{0.5pt}
    \fbox{\includegraphics[width=0.9\dimexpr\textwidth-2\fboxrule-2\fboxsep\relax,page=5, clip]{figures/Appendix_Tab_Tripod_LLM_Checklist.pdf}}
    \vspace{0.1cm}
    \caption{\textbf{TRIPOD-LLM Checklist} (Part 5)}
\end{table}

\begin{table}[ht!]
    \setlength{\fboxsep}{0pt}
    \setlength{\fboxrule}{0.5pt}
    \fbox{\includegraphics[width=0.9\dimexpr\textwidth-2\fboxrule-2\fboxsep\relax,page=6, clip]{figures/Appendix_Tab_Tripod_LLM_Checklist.pdf}}
    \vspace{0.1cm}
    \caption{\textbf{TRIPOD-LLM Checklist} (Part 6)}
\end{table}


\clearpage
\newpage
\setlength\bibitemsep{3pt}
\printbibliography
\clearpage
\end{refsection}

\end{document}